\documentstyle[preprint,prl,aps]{revtex}
\begin{document}
\def\sn2{$\sin^22\theta$}
\def\dm2{$\Delta m^2$}
\def\ch2{$\chi^2$}
\def\ltap{\ \raisebox{-.4ex}{\rlap{$\sim$}} \raisebox{.4ex}{$<$}\ }
\def\gtap{\ \raisebox{-.4ex}{\rlap{$\sim$}} \raisebox{.4ex}{$>$}\ }

\newcommand{\EQ}{\begin{equation}}
\newcommand{\EN}{\end{equation}}
\draft


\def\bec{\begin{center}}
\def\eec{\end{center}}

\def\beq{\begin{equation}}
\def\eeq{\end{equation}}

\def\degres{\mbox{$^\circ$}}      

\def\gsim{\mbox{$\stackrel{_>}{_\sim}$}} 
\def\lsim{\mbox{$\stackrel{_<}{_\sim}$}} 

%
%
\newcommand{\stdfrac}[2]{{{#1}\over{#2}}}
\renewcommand{\frac}[2]{{{\displaystyle #1}\over{\displaystyle #2}}}

\def\eV{\mbox{ eV}}      
\def\keV{\mbox{ keV}}
\def\MeV{\mbox{ MeV}}

\def\BeSv{\mbox{$^{7}$Be}}     
\def\BHt{\mbox{$^{8}$B}}

\def\calA{\mbox{$\cal A$}}     
\def\calB{\mbox{$\cal B$}}
\def\calC{\mbox{$\cal C$}}
\def\calH{\mbox{$\cal H$}}
\def\calM{\mbox{$\cal M$}}
\def\calN{\mbox{$\cal N$}}
\def\calO{\mbox{$\cal O$}}
\def\calR{\mbox{$\cal R$}}
\def\calS{\mbox{$\cal S$}}

\def\hatg{\mbox{${\hat{g}}$}}   
\def\hath{\mbox{${\hat{h}}$}}
\def\hats{\mbox{${\hat{s}}$}}
\def\hatt{\mbox{${\hat{t}}$}}

\def\dms{\mbox{$\Delta m^2$}}     
\def\dMs{\mbox{$\Delta M^2$}}

\def\SdTv{\mbox{$\sin2\theta_V$}}
\def\SdTvS{\mbox{$\sin^2 2\theta_V$}}
\def\STvS{\mbox{$\sin^2 \theta_V$}}
\def\STv{\mbox{$\sin \theta_V$}}

\def\DeltaMs{\mbox{$\Delta m^2$}}
\def\ThetaV{\mbox{$\theta_v$}}
\def\ThetaM{\mbox{$\theta_m$}}

\def\GF{\mbox{$G_F$}}               

                                    %
\def\numt{\mbox{$\nu_{\mu(\tau)}$}} 
\def\nue{\mbox{$\nu_e$}}            
\def\num{\mbox{$\nu_\mu$}}          
\def\nut{\mbox{$\nu_\tau$}}         
\def\nuone{\mbox{$\nu_1$}}          
\def\nutwo{\mbox{$\nu_2$}}          
\def\nuthree{\mbox{$\nu_3$}}        
\def\nus{\mbox{$\nu_s$}}            
\def\REarth{\mbox{$R_\oplus$}}   
\def\MEarth{\mbox{$M_\oplus$}}
\def\BEarth{\mbox{$B_\oplus$}}
\def\RCore{\mbox{$R_{Core}$}}

\def\TYear{\mbox{$T_{y}$}}

\def\RhoR{\mbox{$\rho_R$}}      
\def\rhoR{\mbox{$\rho_R$}}      

\def\Ye{\mbox{$Y_e$}}                 
\def\YeCore{\mbox{$Y_e(Core)$}}       
\def\YeMantle{\mbox{$Y_e(Mantle$}}    

\def\ubar{\mbox{$\bar{u}$}}           

\def\elmass{\mbox{$m_e$}}       

\def\EDms{\mbox{$E_\nu/\Delta m^2$}} 
\def\Enu{\mbox{$E_\nu$}}         
\def\Enucut{\mbox{$E_{\nu,cut}$}}
\def\Te{\mbox{$T_e$}}            
\def\TeTh{\mbox{$T_{e,th}$}}     
\def\Ee{\mbox{$E_e$}}            
\def\EeTh{\mbox{$E_{e,th}$}}     
                                 %

\def\SnuZr{\mbox{$\calS_0$}}        
\def\Snu{\mbox{$\calS$}}            
                                    %

\def\SeZr{\mbox{$\calS_{0}$}}       
\def\Se{\mbox{$\calS$}}             
\def\Ses{\mbox{$\calS^s$}}          
\def\SeD{\mbox{$\calS^D$}}          
\def\SeN{\mbox{$\calS^N$}}          
\def\SeC{\mbox{$\calS^C$}}          
\def\SeM{\mbox{$\calS^D$}}          
                                    %

\def\Re{\mbox{$\calR$}}             
\def\Res{\mbox{$\calR^s$}}          
\def\ReZr{\mbox{$\calR_{0}$}}       
\def\ReZrs{\mbox{$\calR_0^s$}}      
\def\ReZrD{\mbox{$\calR_{0}^D$}}    
\def\ReZrN{\mbox{$\calR_{0}^N$}}    
\def\ReZrC{\mbox{$\calR_{0}^C$}}    
\def\ReZrM{\mbox{$\calR_{0}^M$}}    
                                    %
\def\ReD{\mbox{$\calR^D$}}          
\def\ReN{\mbox{$\calR^N$}}          
\def\ReC{\mbox{$\calR^C$}}          
\def\ReM{\mbox{$\calR^M$}}          
                                    %

\def\Asym{\mbox{$\calA$}}           
\def\AsymP{\mbox{$\calA_P$}}        
\def\AsymPs{\mbox{$\calA_P^s$}}     
\def\AsymPN{\mbox{$\calA_P^N$}}     
\def\AsymPC{\mbox{$\calA_P^C$}}     
\def\AsymPM{\mbox{$\calA_P^M$}}     
\def\AsymS{\mbox{$\calA_{D-N}$}}    
\def\AsymSs{\mbox{$\calA_{D-N}^s$}} 
\def\AsymSN{\mbox{$\calA_{D-N}^N$}} 
\def\AsymSC{\mbox{$\calA_{D-N}^C$}} 
\def\AsymSM{\mbox{$\calA_{D-N}^M$}} 
\def\AsymR{\mbox{$A_{D-N}$}}        
\def\AsymRs{\mbox{$A_{D-N}^s$}}     
\def\AsymRN{\mbox{$A_{D-N}^N$}}     
\def\AsymRC{\mbox{$A_{D-N}^C$}}     
\def\AsymRM{\mbox{$A_{D-N}^M$}}     
\def\AsymRNCM{\mbox{$A_{D-N}^{N,C,M}$}}   
\def\AsymRNM{\mbox{$A_{D-N}^{N(M)}$}}   
                                    %

\def\deltaS{\mbox{$\delta \calS$}}      
\def\deltaSs{\mbox{$\delta \calS^s$}}   
\def\deltaSD{\mbox{$\delta \calS^D$}}   
\def\deltaSN{\mbox{$\delta \calS^N$}}   
\def\deltaSC{\mbox{$\delta \calS^C$}}   
\def\deltaSM{\mbox{$\delta \calS^M$}}   
                                        %

                                        %
\def\Ps{\mbox{${\bar{P}}_\odot$}}       
\def\APTot{\mbox{${\bar{P}}_\oplus$}}   
\def\PTot{\mbox{$P_{\oplus}$}}          
\def\PTots{\mbox{$P^s_\oplus$}}         
\def\PeTw{\mbox{$P_{e2}$}}              
\def\APeTw{\mbox{$<\PeTw>$}}            
\def\APeTws{\mbox{$<\PeTw>^s$}}         
\def\APeTwN{\mbox{$<\PeTw>^{N}$}}       
\def\APeTwC{\mbox{$<\PeTw>^{C}$}}       
\def\APeTwM{\mbox{$<\PeTw>^{M}$}}       
\def\APeTwDC{\mbox{$<\PeTw>^{DC}$}}     

\def\TResid{\mbox{$T_{res}$}}         
\def\TResids{\mbox{$T_{res}^s$}}      
\def\TResidD{\mbox{$T_{res}^D$}}      
\def\TResidN{\mbox{$T_{res}^N$}}      
\def\TResidC{\mbox{$T_{res}^C$}}      
\def\TResidDC{\mbox{$T_{res}^{DC}$}}  
\def\TResidM{\mbox{$T_{res}^M$}}      
                                      %

\def\DAY{\mbox{\em{Day}}}                        
\def\night{\mbox{\em{Night}}}                    
\def\core{\mbox{\em{Core}}}                      
\def\mantle{\mbox{\em{Mantle}}}                  
\def\deepcore{\mbox{\em{Deep - Core}}}           
\def\standard{\mbox{\em{Standard}}}              
\def\hatdelta{\mbox{$\hat{\delta}$}}             
                                                 %

\def\alphaSun{\mbox{$\alpha_\odot$}}   
\def\deltaSun{\mbox{$\delta_\odot$}}   
\def\RSE{\mbox{$R$}}                   
\def\Ellipt{\mbox{$\epsilon_0$}}       
                                       %
\def\hathmid{\mbox{$\hath_{m}$}}       
\def\hathmmax{\mbox{$\hath_{m,max}$}}  
\def\hathmmin{\mbox{$\hath_{m,min}$}}  
                                       %

\def\PeakM{\mbox{$M$}}                 
\def\PeakC{\mbox{$C$}}                 
\def\PeakCI{\mbox{$C_I$}}              
\def\PeakCO{\mbox{$C_O$}}              
                                       %
\def\PeakA{\mbox{$\calA$}}             
\def\PeakHM{\mbox{$\calH^-$}}          
\def\PeakHP{\mbox{$\calH^+$}}          
                                       %

\def\dseedEe{\mbox{$\frac{d\, \sigma_{\nu_e  } (\Te,E_\nu)}{d\,\Te}$}}
                                       %
\def\dsemdEe{\mbox{$\frac{d\, \sigma_{\nu_\mu} (\Te,E_\nu)}{d\,\Te}$}}
                                       %

\def\daynight{D-N}                        
\def\SK{Super - Kamiokande}
\def\maxim{\mbox{max}}                    
\def\FORTRAN{\mbox{\tt{FORTRAN}}}
\def\MATHEMATICA{\mbox{\tt{MATHEMATICA}}}

\def\deg{\degres}  

\def\pp{\mbox{pp}}               
\def\pep{\mbox{pep}}             
\def\CNO{\mbox{CNO}}             

\def\CdTv{\mbox{$\cos 2 \theta_V$}}
\def\CdTM{\mbox{$\cos 2 \theta_m$}}

\def\lambdaDtc{\mbox{$\lambda_D$}}

\def\NA{Nadir angle}

\draft
\begin{titlepage}
\preprint{\vbox{\baselineskip 10pt{
\hbox{Ref. SISSA 53/98/EP}
\hbox{31 July, 1998}}}}
\vskip -0.4cm
\title{ \bf On the Oscillation Length Resonance 
in the Transitions
of Solar and Atmospheric Neutrinos Crossing the Earth Core}
\author{ M. Chizhov $^{a)}$, M. Maris $^{b)}$ and 
S.T. Petcov $^{c,d)}$\footnote{Also at:
Institute of Nuclear Research and Nuclear Energy, Bulgarian Academy
of Sciences, 1784 Sofia, Bulgaria.}}

\address{a) Department of Physics, University of Sofia, 1164 Sofia, Bulgaria}
\address{b) Osservatorio Astronomico di Trieste, I-34113 Trieste, Italy}
\address{c) Scuola Internazionale Superiore di Studi Avanzati, I-34013
Trieste, Italy}
\address{d) Istituto Nazionale di Fizica Nucleare, 
Sezione di Trieste, I-34013
Trieste, Italy}

\maketitle
\begin{abstract}
\begin{minipage}{5in}
\baselineskip 16pt

Assuming two-neutrino mixing takes place in vacuum, 
we study in detail
the conditions under which the 
$\nu_2 \rightarrow \nu_{e}$ and 
$\nu_{\mu} \rightarrow \nu_{e}$ 
($\nu_e \rightarrow \nu_{\mu(\tau)}$) 
transitions in the Earth are strongly
enhanced by the neutrino oscillation length resonance
when the neutrinos cross the Earth core. 
We show, in particular, 
that the neutrino oscillation 
length resonance is operative also in the
$\bar{\nu}_{\mu} \rightarrow \bar{\nu}_{s}$
(or $\nu_{\mu} \rightarrow \nu_{s}$) transitions 
at small mixing angles. The properties of the  
$\nu_2 \rightarrow \nu_{e}$ and 
$\nu_{\mu} \rightarrow \nu_{e}$
($\nu_e \rightarrow \nu_{\mu(\tau)}$) 
transition probabilities in the corresponding 
resonance regions are examined.
Some implications of our results for 
the transitions of solar and atmospheric neutrinos 
traversing the Earth core, relevant for the interpretation
of the results of the solar and atmospheric 
neutrino experiments, are also discussed.

\end{minipage}
\end{abstract}

\end{titlepage}

\newpage

\hsize 16.5truecm
\vsize 24.0truecm
\def\dm{$\Delta m^2$\hskip 0.1cm }
\def\dmsqua{$\Delta m^2$\hskip 0.1cm}
\def\sn{$\sin^2 2\theta$\hskip 0.1cm }
\def\snf{$\sin^2 2\theta$}
\def\trna{$\nu_e \rightarrow \nu_a$}
\def\trnm{$\nu_e \rightarrow \nu_{\mu}$}
\def\trns{$\nu_e \leftrightarrow \nu_s$}
\def\trnat{$\nu_e \leftrightarrow \nu_a$}
\def\trnmt{$\nu_e \leftrightarrow \nu_{\mu}$}
\def\trne{$\nu_e \rightarrow \nu_e$}
\def\trnst{$\nu_e \leftrightarrow \nu_s$}
\def\nue{$\nu_e$\hskip 0.1cm}
\def\numu{$\nu_{\mu}$\hskip 0.1cm}
\def\nutau{$\nu_{\tau}$\hskip 0.1cm}
\baselineskip 21pt 
\font\eightrm=cmr8
\def\aprle{\buildrel < \over {_{\sim}}}
\def\aprge{\buildrel > \over {_{\sim}}}
\renewcommand{\thefootnote}{\arabic{footnote}}
\setcounter{footnote}{0}
\tightenlines

\leftline{\bf 1. Introduction}
\vskip 0.2cm

\indent The $\nu_2 \rightarrow \nu_{e}$ and 
$\nu_{\mu} \rightarrow \nu_{e}$
($\nu_e \rightarrow \nu_{\mu(\tau)}$) transitions/oscillations of 
solar and atmospheric neutrinos in the Earth, 
caused by 
neutrino mixing (with nonzero mass neutrinos) in vacuum 
\footnote{The $\nu_2 \rightarrow \nu_{e}$
transition probability accounts, as is well-known, for the Earth effect 
in the solar neutrino survival probability
in the case of the MSW two-neutrino  
$\nu_e \rightarrow \nu_{\mu(\tau)}$ and
$\nu_e \rightarrow \nu_{s}$ transition solutions of the
solar neutrino problem, $\nu_{s}$ being a sterile neutrino.},
can be strongly amplified by a new type of resonance
which differs from the MSW one and
takes place when the neutrinos traverse the Earth core on the
way to the detector \cite{SP3198}. 
At small mixing angles ($\sin^22\theta \ltap 0.05$),
the maxima due to this resonance 
in the corresponding transition probabilities,
$P(\nu_2 \rightarrow \nu_{e}) \equiv P_{e2}$ and 
$P(\nu_{\mu} \rightarrow \nu_{e})$
($P(\nu_e \rightarrow \nu_{\mu(\tau)})$),
are absolute maxima and dominate in  $P_{e2}$ and 
$P(\nu_{\mu} \rightarrow \nu_{e})$:
the values of the probabilities at these maxima 
in the simplest case of two-neutrino mixing
are considerably larger -
by a factor of $\sim (2.5 - 4.0)$ ($\sim (3.0 - 7.0)$), 
than the values of $P_{e2}$ and 
$P(\nu_{\mu} \rightarrow \nu_{e}) = P(\nu_e \rightarrow \nu_{\mu(\tau)})$
at the local maxima  
associated with the MSW effect 
taking place in the Earth core (mantle).
The enhancement is less dramatic at large mixing angles.
Even at small mixing angles 
the resonance is relatively wide in the neutrino energy 
(or resonance density) - 
it is somewhat wider than the MSW resonance. It also
exhibits strong energy dependence.

  The conditions for existence of the indicated resonance
in the probabilities $P_{e2}$ and 
$P(\nu_{\mu} \rightarrow \nu_{e})$
($P(\nu_e \rightarrow \nu_{\mu(\tau)})$)
include specific constraints on the neutrino oscillation
length in the Earth mantle and in the Earth core \cite{SP3198}.
When satisfied, these conditions ensure that the 
relevant oscillating factors in the 
$\nu_2 \rightarrow \nu_{e}$ and 
$\nu_{\mu} \rightarrow \nu_{e}$
($\nu_e \rightarrow \nu_{\mu(\tau)}$) transition
probabilities are maximal and that this 
produces a resonance maximum in $P_{e2}$ and 
$P(\nu_{\mu} \rightarrow \nu_{e})$ ($P(\nu_e \rightarrow \nu_{\mu(\tau)})$). 
Accordingly, the term ``neutrino oscillation 
length resonance'' or simply 
``oscillation length resonance'' was used 
in \cite{SP3198} to denote the resonance of interest 
\footnote{The oscillation length resonance can produce deep
absolute minima in the probability $P_{e2}$ as well (see further).}. 
In contrast, the MSW effect is, as is well-known,
a resonance in the neutrino mixing. 
There exists a beautiful analogy between the neutrino 
oscillation length resonance and the electron 
spin-flip resonance taking place in a specific configuration
of two magnetic fields (see \cite{SP3198} for a somewhat 
more detailed discussion)
\footnote{This analogy was brought to the attention
of the author of \cite{SP3198} by L. Wolfenstein.}.
The neutrino oscillation length resonance is also 
in many respects similar to the resonance transforming a 
circularly polarized LH photon into a RH photon 
when the photon traverses three layers of optically active 
medium such that the optical activity and the length of the path of the photon
in the first and the third layers are identical but differ from those in the
second layer. 

  The implications of the oscillation length resonance 
enhancement of the probability $P_{e2}$ 
for the tests of the MSW  
$\nu_e \rightarrow \nu_{\mu (\tau)}$ 
and $\nu_e \rightarrow \nu_{s}$ 
transition solutions of the solar neutrino problem 
are discussed in ref. \cite{SP3198} and in much greater 
detail in refs. \cite{Art2,Art3} (see also \cite{Art1}). 
It is quite remarkable that for values of
$\Delta m^2 \cong (4.0 - 8.0) \times 10^{-6}~{\rm eV^2}$ from the small
mixing angle (SMA) MSW solution region
and the geographical latitudes 
at which the Super-Kamiokande, SNO and ICARUS
detectors are located,
the enhancement takes place 
in the $\nu_e \rightarrow \nu_{\mu (\tau)}$ case for values of the 
$^{8}$B neutrino energy lying in the interval
$\sim (5 - 12)~$MeV to which 
these detectors are sensitive. The resonance maximum in
$P_{e2}$ at $\sin^22\theta = 0.01$
for the trajectory with a Nadir angle $h = 23^{\circ}$, for instance,
is located at a resonance density 
$\rho^{res}_{man} \sim 8.0 ~{\rm g/cm^3}$, which 
corresponds to $E \cong 5.3~(10.5)~$MeV if 
$\Delta m^2 = 4.0~(8.0)\times 10^{-6}~{\rm eV^2}$.
Accordingly, at small mixing angles this 
enhancement is predicted to produce \cite{Art2} 
a much bigger - by a factor of $\sim 6$, 
day-night (D-N) asymmetry in the 
Super-Kamiokande sample of solar neutrino events, 
whose night fraction is due to the {\it core-crossing} 
solar neutrinos, in comparison with
the asymmetry determined by using the {\it whole 
night} Super-Kamiokande event sample
\footnote{The possibility of enhancement
of the Earth effect in the Super-Kamiokande sample of events
generated by the core-crossing solar neutrinos was discussed
in ref. \cite{BalzWen94}, where
on the basis of results obtained for few pairs
of values of 
$\Delta m^2$ and $\sin^22\theta$
it was also suggested
that the effect might be measurable even in the case of small
neutrino mixing angle. However,
no estimates of the magnitude of
the core enhancement were given in this article and
the enhancement was interpreted to be due to the
MSW effect taking place in the Earth core.}.
On the basis of these results it was concluded in \cite{Art2} that 
it may be possible to test a rather large part of the
MSW $\nu_e \rightarrow \nu_{\mu (\tau)}$
small mixing angle (SMA) solution region 
in the $\Delta m^2 - \sin^22\theta$
plane by performing selective, i.e., {\it core} 
D-N asymmetry measurements.
The Super-Kamiokande collaboration has already successfully
applied this approach to the analysis of their solar 
neutrino data in terms of the MSW effect 
\cite{SKSuzukinu98,SKNakahata98}:
the limit the collaboration has obtained on the
D-N asymmetry utilizing only the {\it core} 
event sample permitted to 
exclude a part of the SMA solution region allowed by
the mean event rate data from all solar neutrino
experiments (Homestake, GALLEX, SAGE, Kamiokande and Super-Kamiokande).
The SMA solution values of $\sin^22\theta$ and 
$\Delta m^2$ thus excluded belong to the intervals
$\sin^22\theta \cong (0.007 - 0.01)$ and
$\Delta m^2 \cong (5.0\times 10^{-6} - 10^{-5})~{\rm eV^2}$.
In contrast, the current upper limit on 
the D-N asymmetry, derived by the 
Super-Kamiokande collaboration 
using the {\it whole night} data sample of solar neutrino events,
does not permit to probe the SMA solution region: the predicted
asymmetry is too small \cite{Art2}.
The oscillation length resonance leads also
to a relatively large D-N asymmetry
in the recoil - e$^{-}$ spectrum which is being measured 
in the Super-Kamiokande experiment \cite{Art2}.

  As was shown in \cite{SP3198},  
the conditions for the neutrino oscillation length 
resonance are not even approximately
fulfilled if solar neutrinos take part 
in $\nu_e \rightarrow \nu_{s}$ transitions, and hence
the effect of strong enhancement of the probability
$P_{e2}$ does not take place in this case.
However, when the  MSW
resonance occurs in the Earth core, the purely MSW 
enhancement of $P_{e2}$ at small mixing angles
is rather strongly ``assisted'' by 
some of the additional  
interference terms present in $P_{e2}$ (see
ref. \cite{SP3198} for details) and
at the corresponding maximum $P_{e2}$ 
has a noticeably larger value (by a factor of
$\sim (2 - 4)$) than if it were determined by 
the MSW effect only.
Still, at $\sin^22\theta \ltap 0.02$,
the value of $P_{e2}$ at its absolute maximum
in the $\nu_e \rightarrow \nu_{s}$ case is substantially 
smaller (by a factor of $\sim (2.5 - 4.0)$ \cite{SP3198,Art3}) than the
maximum value of $P_{e2}$ 
due to the oscillation length resonance 
if $\nu_e \rightarrow \nu_{\mu (\tau)}$
transitions took place. In addition, due basically to 
the well-known difference 
between the $\nu_{\mu (\tau)}$ and $\nu_{s}$ effective potentials 
in matter and the fact that the Earth matter is approximately
isotopically symmetric, for fixed $\Delta m^2$ and $\sin^22\theta$  
the absolute maximum of $P_{e2}$ occurs in the case of
$\nu_e \rightarrow \nu_{s}$ transitions 
at values of the neutrino energy $E$ exceeding
approximately by a factor 2 the values at 
which the oscillation length resonance maximum takes place
if the solar $\nu_e$ undergo transitions into
$\nu_{\mu (\tau)}$ \cite{Art3}. 
As a consequence, the oscillation length
resonance does not produce such a dramatic enhancement 
of the D-N asymmetry in the {\it core} sample of the
Super-Kamiokande solar neutrino events as in the case of
$\nu_e \rightarrow \nu_{\mu (\tau)}$ transitions.
Thus, the possibility to probe the
MSW $\nu_e \rightarrow \nu_{s}$ (SMA)
solution of the solar neutrino problem via
the measurement of the D-N effect related observables 
in the Super-Kamiokande experiment appears 
unlikely at present \cite{Art3}. 

 The oscillation length resonance enhancement can be present
also in the $\nu_{\mu} \rightarrow \nu_{e}$ and 
$\nu_{e} \rightarrow \nu_{\mu (\tau)}$
transitions of atmospheric neutrinos 
crossing the Earth core \cite{SP3198}.
It can take place practically for all neutrino 
trajectories through the core, in particular,
for the trajectories with $h = (0^{\circ} - 23^{\circ})$. 
In the case of two-neutrino mixing and
at small mixing angles, the maximum in 
$P(\nu_{\mu} \rightarrow \nu_{e}) = 
P(\nu_{e} \rightarrow \nu_{\mu (\tau)})$
due to the oscillation length resonance is an
absolute maximum and dominates in these probabilities
\footnote{Let us note that 
the oscillation length resonance 
enhancement of $P(\nu_{\mu} \rightarrow \nu_{e})$
is operative at large mixing angles as well, at which the probabilities
$P_{e2}$ and $P(\nu_{\mu} \rightarrow \nu_{e})$ exhibit different
dependence on $E/\Delta m^2$ (see further).}. 
This can be especially relevant for the interpretation
of the results of the atmospheric neutrino experiments
and for the future studies of the oscillations/transitions
of atmospheric neutrinos crossing the Earth.
As is well-known, the Super-Kamiokande 
collaboration has reported recently
strong experimental evidences for oscillations
of the atmospheric $\nu_{\mu}$ ($\bar{\nu}_{\mu}$) 
neutrinos \cite{SKKajitanu98,SKAtmosph98}.
Assuming two-neutrino mixing, the data is best described   
in terms of $\nu_{\mu} \leftrightarrow \nu_{\tau}$ and
$\bar{\nu}_{\mu} \leftrightarrow \bar{\nu}_{\tau}$
vacuum oscillations with parameters 
$\Delta m^2 \cong (5\times 10^{-4} - 6\times 10^{-3})~{\rm eV^2}$
and $\sin^22\theta \cong (0.8 - 1.0)$, although 
the hypothesis of large mixing 
$\nu_{\mu} \leftrightarrow \nu_{s}$ oscillations
provides almost equally good 
fit to the data. At the same time, the possibility of 
two-neutrino $\nu_{\mu} \leftrightarrow \nu_{e}$  
and $\bar{\nu}_{\mu} \leftrightarrow \bar{\nu}_{e}$
large mixing oscillations 
is disfavored by the presently existing Super-Kamiokande 
data \cite{SKKajitanu98,SKAtmosph98}. At 
$\Delta m^2 \gtap 2\times 10^{-3}~{\rm eV^2}$
this possibility is ruled out by 
the results of the CHOOZ experiment \cite{CHOOZ98}.

   It is a remarkable coincidence that, 
as was shown in \cite{SP3198},
for values of $\Delta m^2 \sim (5\times 10^{-4} - 5\times 10^{-3})~{\rm eV^2}$
and small mixing angles, $\sin^22\theta \ltap 0.10$,
the neutrino oscillation length resonance 
in $P(\nu_{\mu} \rightarrow \nu_{e})$ 
($P(\nu_{e} \rightarrow \nu_{\mu (\tau)})$)
occurs for values of the energy $E$ of the atmospheric 
$\nu_{e}$ and  $\nu_{\mu}$ 
which contribute either to the 
sub-GeV or to the multi-GeV samples of $e-$like and $\mu-$like events
in the Super-Kamiokande experiment.  
For $\sin^22\theta = 0.01$,
$\Delta m^2 = 5\times 10^{-4};~10^{-3};~5\times 10^{-3}~{\rm eV^2}$, 
and $h = 0^{\circ}$ (Earth center crossing), for instance, 
the absolute maximum in $P(\nu_{\mu} \rightarrow \nu_{e}) = 
P(\nu_{e} \rightarrow \nu_{\mu (\tau)})$ due to the 
oscillation length resonance takes place at
$E \cong 0.75;~1.50;~7.5~$GeV.
Thus, for values of
$\Delta m^2 \sim (5\times 10^{-4} - 6\times 10^{-3})~{\rm eV^2}$
of the region of the $\nu_{\mu} \leftrightarrow \nu_{\tau}$
oscillation solution of the atmospheric neutrino problem, 
the oscillation length resonance strongly enhances the
$\nu_{\mu} \rightarrow \nu_{e}$ (and 
$\nu_{e} \rightarrow \nu_{\mu (\tau)}$)
transitions of the atmospheric neutrinos crossing the Earth core,
making the transition probabilities large
and the transitions perhaps detectable
even at small mixing angles.
Actually, it was suggested in \cite{SP3198,SPNE98} 
that the small excess of e-like events
in the region $-1 \leq \cos\theta_{z}\leq -0.6$,
$\theta_{z}$ being the Zenith angle, either
in the sub-GeV or in the
multi-GeV sample of atmospheric neutrino events, observed  
(in both samples) in the
Super-Kamiokande experiment \cite{SKKajitanu98,SKAtmosph98}, 
is due to $\nu_{\mu} \rightarrow \nu_{e}$
small mixing angle, $\sin^22\theta_{e\mu} \cong (0.01 - 0.10)$, transitions
with $\Delta m^2 \sim (5\times 10^{-4} - 10^{-3})~{\rm eV^2}$ or
respectively $\Delta m^2 \sim (2\times 10^{-3} - 6\times10^{-3})~{\rm eV^2}$, 
strongly enhanced by the neutrino oscillation length
resonance as neutrinos cross the Earth core on the  way
to the detector. The same resonantly enhanced transitions with 
$\Delta m^2 \sim (2\times 10^{-3} - 6\times10^{-3})~{\rm eV^2}$
($\Delta m^2 \sim (5\times 10^{-4} - 10^{-3})~{\rm eV^2}$)
should produce \cite{SP3198,SPNE98} at 
least part of the strong Zenith angle dependence, exhibited by the 
$\mu-$like multi-GeV (sub-GeV) Super-Kamiokande data.

     The $\nu_{\mu} \rightarrow \nu_{e}$ and 
$\nu_{e} \rightarrow \nu_{\mu (\tau)}$
transitions of atmospheric neutrinos 
under discussion should exist if
three-flavour-neutrino (or four-neutrino)  
mixing takes place in vacuum. More specifically, 
such transitions are naturally predicted to take place 
in the case of three-neutrino mixing \cite{SP3198} 
if large mixing angle
$\nu_{e} \leftrightarrow \nu_{\mu}$ oscillations
with $\Delta m^2_{21} \sim 10^{-10}~{\rm eV^2}$
or small mixing angle MSW
$\nu_{e} \rightarrow \nu_{\mu}$ transitions
with $\Delta m^2_{21} \sim (4 - 8)\times 10^{-6}~{\rm eV^2}$,
provide the solution of the solar neutrino
problem, and if the atmospheric neutrino anomaly is
due to $\nu_{\mu} \leftrightarrow \nu_{\tau}$
large mixing angle oscillations with
$\Delta m^2_{31} \sim (5\times 10^{-4} - 6\times 10^{-3})~{\rm eV^2}$.
If $\Delta m^2_{31} \gg \Delta m^2_{21}$, which seems
to be a plausible possibility in view of the 
neutrino oscillation interpretation
of the solar and atmospheric neutrino data, 
the relevant three-neutrino 
$\nu_{\mu} \rightarrow \nu_{e}$ 
and $\nu_{e} \rightarrow \nu_{\mu (\tau)}$
transition probabilities reduce to a two-neutrino
transition probability (see, e.g., \cite{3nuSP88}) 
with $\Delta m^2_{31}$  and 
$\sin^22\theta_{13} \equiv \sin^22\theta_{e\mu} =
4|U_{e3}|^2(1 - |U_{e3}|^2)$ playing 
the role of the relevant
two-neutrino oscillation parameters,
where $U_{e3}$ is the element of the lepton
mixing matrix $U$, coupling the electron and the 
third (heaviest) massive
neutrino, $\nu_3$, in the charged lepton current.
Under the above assumptions and taking 
into account the CHOOZ result \cite{CHOOZ98} one finds that 
$|U_{e3}|^2$ cannot be large, but is also
not necessarily exceedingly small:
$\sin^22\theta_{13} \ltap 0.20$. Thus,
searching for the 
$\nu_{\mu} \rightarrow \nu_{e}$
and $\nu_{e} \rightarrow \nu_{\mu (\tau)}$ 
transitions of atmospheric neutrinos,
amplified by the oscillation length resonance,
can provide unique information, in particular,
about the magnitude of the element 
$U_{e3}$ of the lepton mixing matrix \cite{SPal98}.
As is not difficult to show using the results of \cite{3nuSP88},  
the fluxes of atmospheric $\nu_{e,\mu}$ having energy $E$
and crossing the Earth along a trajectory with Zenith angle $\theta_{z}$
before reaching the detector, $\Phi_{\nu_{e,\mu}}(E,\theta_{z})$, are given 
by the following expressions in the three-neutrino mixing scheme
under discussion \cite{SPNE98}:
  $$\Phi_{\nu_e}(E,\theta_{z}) \cong 
\Phi^{0}_{\nu_e}~\left (  1 + 
  [s^2_{23}r(E,\theta_{z}) - 1]~ 
 P_{2\nu}(E,\theta_{z}; 
 \Delta m^2_{31}, \theta_{13}) \right ),~~\eqno(1a)$$
$$\Phi_{\nu_{\mu}}(E,\theta_{z}) \cong \Phi^{0}_{\nu_{\mu}} ( 1 +
 s^4_{23}~ [(s^2_{23}~r(E,\theta_{z}))^{-1} - 1]~P_{2\nu}(E,\theta_{z}; 
 \Delta m^2_{31}, \theta_{13})~~~~~~~~~~~~~~~~$$ 
$$~~~~~~~~~~~~~~~~-  2c^2_{23}s^2_{23}~\left [ 1 -
Re~( e^{-i\kappa}
A_{2\nu}(\nu_{\tau} \rightarrow \nu_{\tau}))\right ] ),~ \eqno(1b)$$

\noindent Here $\Phi^{0}_{\nu_{e(\mu)}} = 
\Phi^{0}_{\nu_{e(\mu)}}(E,\theta_{z})$ is the
$\nu_{e(\mu)}$ flux in the absence of neutrino transitions/oscillations,
$$s^2_{23} \equiv \frac{|U_{\mu 3}|^2}{1 - |U_{e3}|^2} \leq 1,~~~~~~~~ 
r(E,\theta_{z}) \equiv \frac{\Phi^{0}_{\nu_{\mu}}(E,\theta_{z})}
{\Phi^{0}_{\nu_{e}}(E,\theta_{z})}~,\eqno(1c)$$
\noindent $U_{\mu 3}$ being the $\mu - \nu_3$ element of $U$,
$c^2_{23} = 1 - s^2_{23}$,
$P_{2\nu}(E,\theta_{z};\Delta m^2_{31}, \theta_{13})$ is the probability 
of two-neutrino 
transitions in the Earth which coincides in form
with $P(\nu_{e} \rightarrow \nu_{\tau })$,
$P_{2\nu} = P(\nu_{e} \rightarrow \nu_{\tau})$,
but describes $\nu_{e} \rightarrow \nu_{\tau}$ 
transitions only in a certain limit 
\cite{3nuSP88}, and $\kappa$ and 
$A_{2\nu}(\nu_{\tau} \rightarrow \nu_{\tau})$ are known phase and 
two-neutrino transition probability amplitude
\footnote{Analytic expressions for
$P(\nu_{e} \rightarrow \nu_{\tau})$,
$\kappa$ and $A_{2\nu}(\nu_{\tau} \rightarrow \nu_{\tau})$
will be given later (see eqs. (13), (15) and (16).}. 
The interpretation of the Super-Kamiokande atmospheric 
neutrino data in terms of $\nu_{\mu} \leftrightarrow \nu_{\tau}$
oscillations requires the parameter 
$s^2_{23}$ to lie approximately in the interval
(0.30 - 0.70), with 0.5 being the statistically 
preferred value. For the predicted
ratio $r(E,\theta_{z})$ of the atmospheric 
$\nu_{\mu}$ and $\nu_e$ fluxes 
for the Earth core crossing neutrinos,
$-1 \leq \cos\theta_{z}\ltap -0.8$, one has 
\cite{Honda,Bartol}:
$r(E,\theta_{z}) \cong (2.0 - 2.5)$ for 
the neutrinos giving the 
contribution to the sub-GeV
samples of Super-Kamiokande events, 
and $r(E,\theta_{z}) \cong (2.6 - 4.5)$ for those 
giving the main contribution to the multi-GeV samples.
If $s^2_{23} = 0.5$ and $r(E,\theta_{z}) \cong 2.0$
we have $(s^2_{23}~r(E,\theta_{z}) - 1) \cong 0$ and
the possible effects of the 
$\nu_{\mu} \rightarrow \nu_{e}$ 
and $\nu_{e} \rightarrow \nu_{\mu (\tau)}$ 
transitions on the $\nu_e$ flux and correspondingly on 
the sub-GeV $e-$like sample of events, 
would be strongly suppressed even 
if these transitions are maximally
enhanced by the oscillation length resonance.
The indicated suppression may actually be taking place
\footnote{Indeed, a more detailed investigation \cite{SPal98}
performed within the indicated three-neutrino mixing scheme
reveals, in particular, that the excess 
of e-like events in the Super-Kamiokande sub-GeV data 
at $-1 \leq \cos\theta_{z}\leq -0.6$
seems unlikely to be due to small mixing angle,
$\sin^22\theta_{13} \ltap 0.20$, $\nu_{\mu} \rightarrow \nu_{e}$
transitions amplified by the oscillation length resonance.}.
However, one should keep in mind that the factor
$(s^2_{23}~r(E,\theta_{z}) - 1)$ can be as large as
$\sim 0.6$ for the sub-GeV neutrinos and that 
the effects of the resonantly enhanced
$\nu_{\mu} \rightarrow \nu_{e}$ and 
$\nu_{e} \rightarrow \nu_{\mu (\tau)}$ 
transitions in the sub-GeV $e-$like event sample 
can be suppressed by the specific way this sample is 
selected from the data in 
the Super-Kamiokande experiment \cite{SPal98}. 
For the multi-GeV neutrinos we have 
$(s^2_{23}~r(E,\theta_{z}) - 1) \gtap 0.3~(0.9)$ for 
$s^2_{23} = 0.5~(0.7)$.  
The factor $(s^2_{23}~r(E,\theta_{z}) - 1)$ 
amplifies the effect of the 
$\nu_{\mu} \rightarrow \nu_{e}$ 
transitions in the $e-$like sample 
for $E \gtap (5 - 6)~{\rm GeV}$, for which
$r(E,\theta_{z}) \gtap 4$ \cite{Honda,Bartol}. 
This discussion suggests that 
the effects of the neutrino 
oscillation length resonance
in the Super-Kamiokande $e-$like multi-GeV data 
for $\Delta m^2_{31} \sim (2\times 10^{-3} - 6\times10^{-3})~{\rm eV^2}$ 
can be much larger than those  
in the $e-$like sub-GeV data
for $\Delta m^2_{31} \sim (5\times 10^{-4} - 
10^{-3})~{\rm eV^2}$ \cite{SPNE98,SPal98}. 
As similar analysis of the expression (1b) for 
$\Phi_{\nu_{\mu}}$ shows,
they are also expected to be larger
(within the three-neutrino mixing scheme considered)
than the oscillation length resonance effects in the
multi-GeV $\mu-$like event sample \cite{SPNE98,SPal98}. 

   In view of the possible important 
phenomenological implications of the
neutrino oscillation length resonance, some of which 
have been discussed above, we continue in this article the 
studies of the resonance in the $\nu_{2} \rightarrow \nu_{e}$ 
and $\nu_{\mu} \rightarrow \nu_{e}$
($\nu_{e} \rightarrow \nu_{\mu (\tau)}$) transitions,
began in ref. \cite{SP3198}.
Assuming two-neutrino $\nu_e - \nu_{\mu (\tau)}$ or 
$\nu_e - \nu_{s}$ mixing (with nonzero-mass neutrinos) 
exists in vacuum, we study 
in greater detail the conditions under which
the neutrino oscillation length resonance 
takes place in the $\nu_{2} \rightarrow \nu_{e}$ 
and $\nu_{\mu} \rightarrow \nu_{e}$
($\nu_{e} \rightarrow \nu_{\mu (\tau)}$) transitions
of the solar and atmospheric neutrinos in the Earth.
We give a graphical representation of both
the necessary conditions for existence of the 
extrema in the probabilities 
$P_{e2}$ and $P(\nu_{\mu} \rightarrow \nu_{e})$ 
($P(\nu_{e} \rightarrow \nu_{\mu (\tau)}$), 
associated with the resonance,
and of the supplementary conditions ensuring
that these extrema correspond to maxima. This permits, e.g., 
to determine all relevant values of the 
transition parameters, $E$, $\Delta m^2$ and
$\sin^22\theta$, at which the resonance 
occurs for a given neutrino trajectory through the Earth core.
Similar analysis is performed for the case of 
$\bar{\nu}_{\mu} \rightarrow \bar{\nu}_{s}$
transitions at small mixing angles.
We also investigate the behavior of the probabilities 
$P_{e2}$ and $P(\nu_{\mu} \rightarrow \nu_{e})$ 
($P(\nu_{e} \rightarrow \nu_{\mu (\tau)}$) in the region of the
neutrino oscillation length resonance.
We show, in particular, that at small mixing angles 
the resonance is relatively wide not only
in the variable $E/\Delta m^2$, but also in the
Nadir angle variable $h$.
The implications of our results
for the current and future experiments with solar 
and atmospheric 
neutrinos are also briefly discussed.

   Let us note that the Earth enhancement of the two-neutrino
$\nu_2 \rightarrow \nu_{e}$ and
$\nu_{\mu (e)} \rightarrow \nu_{e (\mu)}$
($\nu_e \rightarrow \nu_{\mu(\tau)}$)
transitions of interest of the solar 
and atmospheric neutrinos
in the case of 
relatively small mixing angles
has been discussed 
in the past in, e.g., 
refs. \cite{Art1,BalzWen94,PastAtmo,3nuKP88,PastSun,Rosen97}.
Some of these articles contain plots
of the probabilities $P_{e2}$ and/or
$P(\nu_{\mu} \rightarrow \nu_{e})$ or
$P(\nu_e \rightarrow \nu_{\mu(\tau)})$
on which one can clearly
recognize now the dominating maximum due to the neutrino
oscillation length resonance (see, e.g., 
\cite{PastAtmo,3nuKP88,PastSun,Rosen97}).
However, this maximum was invariably interpreted
to be due to the MSW effect in the Earth core
before the appearance of the study \cite{SP3198}.

\vskip 0.3cm
\leftline{\bf 2. The Earth Model, the Two-Layer Density Approximation and}

\hskip 0.1truecm {\bf the Neutrino Oscillation Length Resonance}
\vglue 0.2cm

     We begin with a brief summary of the main properties of the reference
Earth model and of the results obtained in 
ref. \cite{SP3198}, which will be utilized in our 
further analysis. Following \cite{SP3198,Art2,Art3,Art1,3nuKP88} we 
will use the Stacey 
model from 1977 \cite{Stacey:1977} as a 
reference Earth model. The density distribution
in the Stacey model is,
as in all Earth models known to us, spherically symmetric
and there are two major density structures - 
the core and the mantle, and 
a certain number of substructures (shells or layers).
The core has a radius $R_c = 3485.7~$km,
the Earth mantle depth is approximately $R_{man} = 2885.3~$km,
and the Earth radius in the Stacey model 
is $R_{\oplus} = 6371~$km.  
The mean values of the matter densities of the core and of the mantle 
read, respectively: $\bar{\rho}_c \cong 11.5~{\rm g/cm^3}$ and 
$\bar{\rho}_{man} \cong 4.5~{\rm g/cm^3}$.
The density distribution in the 1977 Stacey model
practically coincides with that 
in the more recent PREM model \cite{PREM81}.

  All the interesting features of the 
solar and atmospheric neutrino transitions 
in the Earth, including those related to 
the neutrino oscillation length resonance,
can be understood quantitatively 
in the framework of the two-layer model of the Earth density
distribution \cite{SP3198}.
The density profile of the Earth in the two-layer model 
is assumed to consist of two structures - 
the mantle and the core, 
having different densities, $\rho_{man}$ and $\rho_{c}$, and different
electron fraction numbers, $Y_e^{man}$ and $Y_e^{c}$, none of which however 
vary within a given structure. The core radius and the depth of the mantle
are known with a rather good precision and these data are incorporated
in the Earth models \cite{Stacey:1977,PREM81}.
The densities $\rho_{man}$ and $\rho_{c}$
in the case of interest should be considered as mean effective densities
along the neutrino trajectories, which can vary somewhat with
the change of the trajectory
\footnote{Let us note that 
because of the spherical symmetry of the Earth, 
a given neutrino trajectory through the Earth is completely 
specified by its Nadir angle (see, e.g., \cite{Art1}).}: 
$\rho_{man} = \bar{\rho}_{man}$ and 
$\rho_c = \bar{\rho}_{c}$.
In the Stacey model one has:
$\bar{\rho}_{man} \cong (4 - 5)~ {\rm g/cm^3}$ and 
$\bar{\rho}_{c} \cong (11 - 12)~ {\rm g/cm^3}$.
For the electron fraction numbers in the mantle and in the core one can use
the standard values \cite{Stacey:1977,PREM81,CORE} 
(see also \cite{Art2}) $Y_e^{man} = 0.49$ and $Y_e^{c} = 0.467$.
Numerical calculations show 
\cite{MP98:2layers} (see also \cite{3nuKP88}) that, e.g., the 
(time-averaged probability $P_{e2}$ calculated within 
the two-layer model of the Earth with $\bar{\rho}_{man}$
and $\bar{\rho}_{c}$ taken from
the Stacey 1977 model \cite{Stacey:1977} 
reproduces with a remarkably 
high precision the probability calculated 
by solving numerically the relevant system of evolution equations
with the much more sophisticated Earth density profile
of the Stacey model \cite{Stacey:1977}. 

  The probability of the $\nu_2 \rightarrow \nu_{e}$ 
transition in the Earth, $P_{e2}$,
which accounts for the Earth effect 
in the solar neutrino survival probability,
has the following form in the two-layer model 
of the Earth density distribution \cite{SP3198}:
$$P_{e2} = \sin^2\theta~ +~ {1\over {2}} 
\left [1 - \cos \Delta E''X'' \right ] \left [ \sin^2(2\theta''_{m} - 
 \theta) -  \sin^2\theta \right ]~~~~~~~~~~~~~~~~~~~~~~~~~~~~~~~~~~~~~~~~~$$
$$ +~ {1\over {4}} \left [1 - \cos \Delta E''X'' \right ]
\left [1 - \cos \Delta E'X' \right ] 
\left [ \sin^2(2\theta''_{m} - 4\theta'_{m}  + \theta) - 
 \sin^2(2\theta''_{m} - \theta) \right ]~~~~~~~~~~ $$
$$ - ~{1\over {4}} \left [1 - \cos \Delta E''X'' \right ]
\left [1 - \cos 2\Delta E'X' \right ] 
\left [ \sin^2(2\theta'_{m} - \theta) - 
 \sin^2\theta \right ] \cos^2(2\theta''_{m} - 2\theta'_{m})~~~~~~ $$
$$~~+~{1\over {4}} \left [1 + \cos \Delta E''X'' \right ]
\left [1 - \cos 2\Delta E'X' \right ] 
\left [ \sin^2 (2\theta'_{m} - \theta) - 
 \sin^2\theta \right ]~~~~~~~~~~~~~~~~~~~~~~~~~~~~~ $$
$$ +~{1\over {2}} \sin \Delta E''X'' ~\sin 2\Delta E'X'  
\left [ \sin^2(2\theta'_{m} - \theta) - 
 \sin^2\theta \right ] \cos(2\theta''_{m} - 2\theta'_{m})
~~~~~~~~~~~~~~~~~~~~ $$
$$~~~~~+~ {1\over {4}} \left [ \cos (\Delta E'X' - \Delta E''X'') 
- \cos (\Delta E'X' + \Delta E''X'') \right ]
 \sin (4\theta'_{m} - 2\theta)  
 \sin (2\theta''_{m} - 2\theta'_m).~ \eqno(2)$$  

\noindent Here
$$\Delta E'~(\Delta E'') = 
\frac{\Delta m^2}{2E}
\sqrt{\left( 1 - \frac{\bar{\rho}_{man~(c)}}{\rho^{res}_{man~(c)}}
\right)^2 \cos^22\theta 
+ \sin^22\theta }~,~
\eqno(3)$$
\noindent is the difference between the energies of the two energy-
eigenstate neutrinos in the mantle (core),
$\theta'_{m}$ and  $\theta''_{m}$ are 
the mixing angles in matter in the
mantle and in the core, respectively,

$$\sin^22\theta'_{m}~(\sin^22\theta''_{m}) = \frac{\sin^22\theta}
{(1 - \frac{\bar{\rho}_{man(c)}}{\rho^{res}_{man(c)}})^2 \cos^22\theta + 
\sin^22\theta },~~~\eqno(4)$$

\noindent $X'$ is half of the distance the neutrino
travels in the mantle and $X''$ is the length of 
the path of the neutrino in the core,
and $\rho^{res}_{man}$ and $\rho^{res}_{c}$
are the resonance densities in the mantle and in the
core. The latter can be obtained from the expressions 
$$\rho^{res} = \frac{\Delta m^2 \cos2\theta}{2E\sqrt{2} G_F f(Y_e)}~m_{N},~
~~~~~~~\eqno(5)$$ 
$$f(Y_e) = Y_e,~~~\nu_e \rightarrow \nu_{\mu (\tau)}~
(\nu_{\mu} \rightarrow \nu_{e})~transitions,~~~~\eqno (6a)$$
$$f(Y_e) = \frac{1}{2}(3Y_e - 1),~~~\nu_e \rightarrow \nu_{s}~
(\nu_s \rightarrow \nu_{e})~transitions,~~~~
 \eqno (6b)$$
\noindent $m_{N}$ being the nucleon mass,
by using the specific values of $Y_e$ in the mantle and in the core.
We have $\rho^{res}_{man}\neq \rho^{res}_{c}$  (but
$\rho^{res}_{man} \sim \rho^{res}_{c}$)
because $Y_e^{c} = 0.467$ and $Y_e^{man} = 0.49$.
Obviously, 
one has: $f(Y_e^{c})\rho^{res}_{c} = f(Y_e^{man})\rho^{res}_{man}$.

  We will present some of our results 
(obtained for fixed $\sin^22\theta$) utilizing the density parameter 
$$\rho_r = \frac{\Delta m^2 \cos 2\theta}
{2 E \sqrt{2} G_{F} C_{a(s)}}m_{N}~,~~\eqno(7a)$$
\vskip 0.3truecm
\noindent  where the constant $C_{a(s)} = 0.50~(0.25)$ for the
$\nu_e \rightarrow \nu_{\mu (\tau)}$
($\nu_e \rightarrow \nu_{s}$) transitions.
The parameter $\rho_r$ would coincide with the
resonance density if $Y_e$ were equal to
$1/2$ both in the mantle and in the core, so 
$\rho_r \cong \rho^{res}_{man}$; for given 
$\sin^22\theta$ it is equivalent to
$E/\Delta m^2$, but gives an idea about the densities
at which one has an enhancement of the
transition probabilities of interest
when the neutrinos cross the Earth and we have used
it for the latter purpose in \cite{Art2,Art3,Art1} and in \cite{SP3198}.
Clearly, $C_a \rho_r = Y_e^{c}\rho^{res}_{c} = Y_e^{man}\rho^{res}_{man}$,
or $C_s \rho_r = 0.5(3Y_e^{c} - 1)\rho^{res}_{c} = 
0.5(3Y_e^{man} - 1)\rho^{res}_{man}$ depending on whether
$\nu_e - \nu_{\mu (\tau)}$ or $\nu_e - \nu_{s}$ 
mixing takes place. The relation between $\rho_r$ and 
$\Delta m^2/E$, expressed in units of ${\rm g/cm^3}$ and 
${\rm eV^2/MeV}$, respectively, has the form:
$$\rho_{r} [g/cm^3] = 1.32\times 10^{7}~\frac{0.5}{C_{a(s)}}~ 
\frac{\Delta m^2 [eV^2]}{E[MeV]}~ cos2\theta.
~~\eqno(7b)$$
\vskip 0.3truecm
  For a neutrino trajectory which is specified by a given Nadir angle $h$ we have:
$$X' = R_{\oplus}\cosh -  \sqrt{R^{2}_{c} - R^2_{\oplus}\sin^2h},~~
X'' = 2\sqrt{R^{2}_{c} - R^2_{\oplus}\sin^2h}.~~\eqno(8)$$
 
 The conditions for the presence of a neutrino oscillation length
resonance maximum in $P_{e2}$ read \cite{SP3198}:
$$\Delta E'X' = \pi (2k + 1),~~ \Delta E''X'' = \pi (2k' + 1), 
~~k,k' = 0,1,2,...~,~\eqno(9)$$
\noindent and 
$$\sin^2(2\theta''_{m} - 4\theta'_{m} + \theta) - \sin^2\theta > 0,
~~\eqno(10a)$$
$$\sin^2(2\theta''_{m} - 4\theta'_{m}  + \theta)\sin(2\theta''_{m} - 
2\theta'_{m})
\sin(2\theta'_{m} - \theta)\cos(2\theta''_{m} - 4\theta'_{m} + \theta)$$
$$ +~\frac{1}{4}\sin^2\theta \sin(4\theta''_{m} - 8\theta'_{m} + 2\theta)
\sin(4\theta''_{m} - 4\theta'_{m}) < 0.~~\eqno(10b)$$

\noindent When equalities (9) hold, conditions (10a) and (10b) ensure that 
$P_{e2}$ has a maximum
\footnote{Note that the factor 1/4 in the second term in the left-hand side of 
the inequality (10b) is missing in the same inequality given in \cite{SP3198}
(see eq. (12b) in \cite{SP3198}).}. At the resonance maximum 
the probability $P_{e2}$ takes the value \cite{SP3198}:
$$P^{max}_{e2} = \sin^2(2\theta''_{m} - 4\theta'_{m}  + \theta).~\eqno(11)$$

\noindent  If instead of inequality (10a) we have 
$$\sin^2(2\theta''_{m} - 4\theta'_{m} + \theta) - \sin^2\theta < 0,
~~\eqno(12)$$

\noindent and inequality (10b) is fulfilled (together with equalities (9)),
the neutrino oscillation resonance extrema in $P_{e2}$ correspond to minima.

    As was already indicated in the Introduction, 
the oscillation length resonance is present in the 
$\nu_{\mu} \rightarrow \nu_{e}$ (and $\nu_{e} \rightarrow \nu_{\mu (\tau)}$) 
transitions of atmospheric neutrinos as well \cite{SP3198}.
The probability of the $\nu_{\mu} \rightarrow \nu_{e}$ 
($\nu_{e} \rightarrow \nu_{\mu (\tau)}$) transitions,
$P(\nu_{\mu} \rightarrow \nu_{e}) = P(\nu_{e} \rightarrow \nu_{\mu (\tau)})$,
can be obtained
\footnote{All subsequent results for the 
$\nu_{\mu} \rightarrow \nu_{e}$ and 
$\nu_{e} \rightarrow \nu_{\mu (\tau)}$ transitions will be formulated for the 
probability $P(\nu_{\mu} \rightarrow \nu_{e})$.}
from eq. (1) by formally
setting $\theta = 0$ while keeping $\theta_{m}' \neq 0$ and 
$\theta_{m}'' \neq 0$:
$$P(\nu_{\mu} \rightarrow \nu_{e}) = P(\nu_{e} \rightarrow \nu_{\mu (\tau)}) 
=  {1\over {2}} 
\left [1 - \cos \Delta E''X'' \right ] \sin^2 2\theta''_{m} ~~~~~~~~~~
~~~~~~~~~~~~~~~~~~~~~~~~~~~~~~~~$$
$$~~~~ +~ {1\over {4}} \left [1 - \cos \Delta E''X'' \right ]
\left [1 - \cos \Delta E'X' \right ] 
\left [ \sin^2(2\theta''_{m} - 4\theta'_{m}) - 
 \sin^2 2\theta''_{m} \right ]~~~~~~~~~~~~~$$
$$~~~~ - ~{1\over {4}} \left [1 - \cos \Delta E''X'' \right ]
\left [1 - \cos 2\Delta E'X' \right ] \sin^2 2\theta'_{m}  
\cos^2(2\theta''_{m} - 2\theta'_{m})~~~~~~~~~~~~~~~~~$$
$$~+~{1\over {4}} \left [1 + \cos \Delta E''X'' \right ]
\left [1 - \cos 2\Delta E'X' \right ]~\sin^2 2\theta'_{m}~~~~~~~~
~~~~~~~~~~~~~~~~~~~~~~~~~~~ $$
$$+ ~{1\over {2}} \sin \Delta E''X'' ~\sin 2\Delta E'X'  
~\sin^2 2\theta'_{m} \cos(2\theta''_{m} - 2\theta'_{m})
~~~~~~~~~~~~~~~~~~~~~~~~~~$$
$$~~~~~ +~{1\over {4}} \left [ \cos (\Delta E'X' - \Delta E''X'') 
- \cos (\Delta E'X' + \Delta E''X'') \right ]
 \sin 4\theta'_{m} 
 \sin (2\theta''_{m} - 2\theta'_m).\eqno(13)$$  

\noindent Correspondingly, 
the maximum conditions (9) are the same as for $P_{e2}$,
while the analogs of the conditions (10a) and (10b) for this
probability change:
condition (10a) becomes
$\sin^2(2\theta''_{m} - 4\theta'_{m}) > 0$ and is always fulfilled,
while condition (10b) transforms into \cite{SP3198}
$$\cos(2\theta''_{m} - 4\theta'_{m}) < 0.~~\eqno(14)$$

\noindent Obviously, in contrast to $P_{e2}$, the probability
$P(\nu_{\mu} \rightarrow \nu_{e})$ cannot have minima associated
with conditions (9). 

  At small mixing angles, $\sin^22\theta < 0.10$,
the maxima in the probabilities 
$P_{e2}$ and $P(\nu_{\mu} \rightarrow \nu_{e})$ 
associated with
the conditions (9) and (10a), (10b), and (9) 
and (14), are absolute maxima.
For given $\sin^22\theta$ and trajectory crossing the Earth core,
the neutrino oscillation length resonance is relatively
wide in the variable $E/\Delta m^2$ (or $\rho_r$, 
see Figs. 1 and 2 in \cite{SP3198}).

   The analytic expression and the resonance
conditions (13) and (14) derived for the probability 
$P(\nu_{\mu} \rightarrow \nu_{e})$
are valid, with minor modifications, for the probabilities
of the i) $\nu_{e} \rightarrow \nu_{s}$ 
($\bar{\nu}_{e} \rightarrow \bar{\nu}_{s}$) and 
ii) $\nu_{\mu} \rightarrow \nu_{s}$ 
($\bar{\nu}_{\mu} \rightarrow \bar{\nu}_{s}$)
transitions, $P(\nu_{e} \rightarrow \nu_{s})$
($P(\bar{\nu}_{e} \rightarrow \bar{\nu}_{s})$) and 
$P(\nu_{\mu} \rightarrow \nu_{s})$ 
($P(\bar{\nu}_{\mu} \rightarrow \bar{\nu}_{s})$):
one has to use in eqs. (3), (4), (9) and (14) and
in the formula for $P(\nu_{\mu} \rightarrow \nu_{e})$, 
the expressions for the resonance densities 
$\rho^{res}_{man,c}$ with 
i) $f(Y_e)$ given by  eq. (6) ($f(Y_e) = 0.5(1 - 3Y_e)$) and ii) 
$f(Y_e) = 0.5(Y_e - 1)$ ($f(Y_e)= 0.5(1 - Y_e)$), respectively.
Since $Y_e$ for the Earth is close to 0.5,
only the probabilities
$P(\nu_{e} \rightarrow \nu_{s})$ and 
$P(\bar{\nu}_{\mu} \rightarrow \bar{\nu}_{s})$ 
can, in principle, be amplified for $\cos2\theta > 0$ 
by the oscillation length
resonance if $\Delta m^2 >  0$, which is assumed throughout this study.
Note, however, that in the case of two-neutrino mixing,
the probability $P(\nu_{e} \rightarrow \nu_{s})$
($P(\bar{\nu}_{\mu} \rightarrow \bar{\nu}_{s})$)
for $\Delta m^2 >  0$ coincides, as can be easily shown, 
with the probability
$P(\bar{\nu}_{e} \rightarrow \bar{\nu}_{s})$
($P(\nu_{\mu} \rightarrow \nu_{s})$) for $\Delta m^2 < 0$ for any
given values of $E$, $|\Delta m^2|$ and $\theta$, and that 
conditions (9) and (14) are identical in the two cases.

  The phase $\kappa$ and the probability 
amplitude $A_{2\nu}(\nu_{\tau} \rightarrow \nu_{\tau})$,
which appear in eq. (1b) for the flux of atmospheric $\nu_{\mu}$
in the case of three-flavour neutrino mixing and strong hierarchy
between the neutrino mass squared differences and therefore 
can play important role in the interpretation of the, 
e.g., Super-Kamiokande
atmospheric neutrino data, have the following form 
in the two-layer model of the Earth density distribution 
\cite{SP3198,SPNE98,3nuSP88}:
$$\kappa \cong {1\over {2}} [ {\Delta m^2_{31}\over{2E}}~X +
\sqrt{2} G_F {1\over{m_N}}(X''Y_e^{c}\bar{\rho}_{c} +
2X'Y_e^{man}\bar{\rho}_{man}) -
2\Delta E'X' - \Delta E''X'']
- {\Delta m^2_{21}\over{2E}}X\cos2\varphi_{12},\eqno(15)$$
$$A_{2\nu}(\nu_{\tau} \rightarrow \nu_{\tau}) = 1 +~
\left( e^{-i2\Delta E'X'} - 1 \right) \left[ 1 +
\left(e^{-i\Delta E''X''} - 1 \right)
\cos^2(\theta'_{m} - \theta''_{m}) \right ]
\cos^2\theta'_{m}~$$

\vspace*{-0.9cm}

$$ +~ \left( e^{-i\Delta E''X''} - 1 \right)
\cos^2\theta''_{m}~+~ {1\over {2}} ~\left( e^{-i\Delta E''X''} - 1 \right )~
\left( e^{-i\Delta E'X'} - 1 \right) \sin(2\theta'_{m} - 2\theta''_{m})
\sin2\theta'_{m}. ~~~\eqno(16)$$
\noindent where $X = X'' + 2X'$ and
\footnote{The expression for 
$A_{2\nu}(\nu_{\tau} \rightarrow \nu_{\tau})$ can be obtained 
from eq. (1) in ref. \cite{SP3198} by formally setting 
$\theta = \pi/2$ while keeping $\theta'_m$ and $\theta''_m$ 
arbitrary, and then interchanging $\sin\theta'_m$ ($\sin \theta''_m$) and
$\cos\theta'_m$ ($\cos\theta''_m$).}
$\cos 2\varphi_{12} = (|U_{e1}|^2 - |U_{e2}|^2)/(1 - |U_{e3}|^2)$.

  Let us note finally that the neutrino oscillation 
length resonance 
arises when the conditions specified above are fulfilled, 
due basically to the interference between 
the probability amplitudes describing the  
neutrino transitions in the Earth core and in the Earth mantle.

\vglue 0.3cm
\leftline{\bf 3. The  Oscillation Length Resonance Conditions and the}

\hskip 0.1truecm {\bf Probabilities $P_{e2}$,~ 
$P(\nu_{\mu} \rightarrow \nu_{e})$ 
($P(\nu_{e} \rightarrow \nu_{\mu (\tau)})$) and 
$P(\bar{\nu}_{\mu} \rightarrow \bar{\nu}_{s})$}

\vglue 0.3cm
\leftline{\bf 3.1. The Resonance Conditions}
\vglue 0.2cm

 The quantities $\Delta E'X'$ and $\Delta E''X''$ represent the phase 
differences the states of the energy-eigenstate neutrinos
acquire after the neutrinos have crossed the mantle 
and the core, respectively. Thus, the neutrino osacillation length resonance
can take place only when these phases are correlated, 
being odd multiples of $\pi$, eq. (9). Since 
$$\Delta E'~(\Delta E'') = \frac{2\pi}{L^{m}_{man(c)}}~,~~\eqno(17)$$

\noindent where $L^{m}_{man(c)}$ is the neutrino oscillation length 
in matter in the Earth mantle (core), and $X'$ and $X''$ are fixed 
for a given neutrino trajectory, conditions (9) are constraints
on the the oscillation lengths  $L^{m}_{man}$ and  $L^{m}_{c}$.
When these constraints are satisfied, the oscillating factors
in the first two terms in the expressions (2) and (13) for 
$P_{e2}$ and $P(\nu_{\mu} \rightarrow \nu_{e})$ 
are maximized. If in addition inequalities (10a) and (10b) or (14)
hold, this produces 
the resonance maximum of interest 
in $P_{e2}$ or $P(\nu_{\mu} \rightarrow \nu_{e})$.  
For these {\it two} reasons 
the term ``neutrino oscillation length resonance''
was used in \cite{SP3198} to denote the effect under discussion.
Although this term may not be perfect,
it underlines the physical essence of the effect.

 Conditions (9) and conditions (10a) and (10b) are shown graphically 
in Figs. 1a - 1f and in Figs. 2a - 2b, respectively, while condition (14) 
is shown in Figs. 3a - 3b. 
Figures 1a - 1f, 2a - 2b and 3a - 3b correspond to the case of 
$\nu_e - \nu_{\mu (\tau)}$ mixing.
We will discuss first conditions (10a), (10b) and (14).
As was noticed in \cite{SP3198}, at small mixing angles, 
$\sin^22\theta \ltap 0.10$,
(10a) is fulfilled for
$$\rho^{res}_{man} < (2r - 1)\bar{\rho}_{man},~~~
\rho^{res}_{man} \neq \bar{\rho}_{man},~~~\eqno(18a)$$ 

\noindent where 
$$r \equiv \frac{\bar{\rho}_{c}~Y_e^{c}}{\bar{\rho}_{man}Y_e^{man}}~
.~~~\eqno(18b)$$
\vskip 0.2truecm
\noindent For the standard values of $\bar{\rho}_{man,c}$ and $Y_e^{man,c}$
given earlier one has $r \cong 2.44$ 
and (18a) corresponds to 
$\rho^{res}_{man} <  17.4~{\rm g/cm^3}$, which is in 
very good agreement with our numerical results (Fig. 2a). 
Actually, as Figs. 2a - 2b show, (10a) is fulfilled 
for any $\theta$ in a subregion of 
the region determined by (18a).  
Condition (10b) holds in two practically 
disconnected regions (Figs. 2a - 2b).
In the region where (10a) is satisfied, condition (10b) reduces to 
the simple inequality \cite{SP3198}:
$$\cos (2\theta''_{m} - 4\theta'_{m}  + \theta) < 0.~~~\eqno(18c)$$  

\noindent Thus, condition (18c) determines effectively the region where
both inequalities (10a) and (10b) are valid at $\sin^22theta \ltap 0.10$. 
Condition (18c) can be satisfied for \cite{SP3198}
$$\bar{\rho}_{man} < \rho^{res}_{man,c} < \bar{\rho}_{c}~,~~\eqno(19a)$$
\noindent as well as for 
$$ 0 < \rho^{res}_{man} < \bar{\rho}_{m}.~~\eqno(19b)$$

\noindent Only the first region, (19a), is relevant for the transitions 
of solar neutrinos with parameters corresponding to the 
small mixing angle  
MSW solution of the solar neutrino problem \cite{SP3198}.
At large mixing angles the oscillation length
resonance enhancement of $P_{e2}$ can take place only 
in the second region, (19b) (see further).

   In the region $\rho^{res}_{man} > (2r - 1)\bar{\rho}_{man}$,
where inequality (12) holds, condition (10b) is satisfied 
at small mixing angles for
$$\frac{2r}{3(r - 1)}
\left ( \sqrt{1 + \frac{4r}{9(r - 1)^2}} - 1 \right )^{-1}~
\bar{\rho}_{man} < \rho^{res}_{man} <  
\frac{3r - 1}{3 - r}~\bar{\rho}_{man}
.~\eqno(20)$$
\vskip 0.2truecm
\noindent For $r \cong 2.44$, 
(20) gives $21.7~{\rm g/cm^3} < \rho^{res}_{man} < 50.3~{\rm g/cm^3}$  - 
in very nice agreement with our numerical results (Fig. 2a). 
Note that, as it follows from Fig. 2a,
condition (10b) is fulfilled in most of the region
determined by (20) even at relatively 
large mixing angles.   

 Condition (14) holds at small mixing angles for (Fig. 3a)   
$$\rho^{res}_{man,c} < \bar{\rho}_{c}, ~~~~
\rho^{res}_{man} \neq \bar{\rho}_{man}~.~~~~\eqno(21)$$

\noindent The regions determined by the two inequalities (10a), (10b) 
and by the inequality (14) differ 
at large mixing angles.

  At small mixing angles, $\sin^22\theta \ltap 0.10$,
conditions (9) can be fulfilled only for  $k = k' = 0 \cite{SP3198}$.  
For the $\nu_2 \rightarrow \nu_{e}$ transitions
in the case of $\nu_e - \nu_{\mu (\tau)}$ mixing, and for the
$\nu_{\mu} \rightarrow \nu_{e}$ and
$\nu_e \rightarrow \nu_{\mu (\tau)}$ 
transitions, taking place along  
the trajectories with $h = 0^{\circ};~13^{\circ};~23^{\circ}$,
one has $\Delta E'X' = \pi$ and $\Delta E''X'' = \pi$
at $\sin^22\theta \cong 0.045;~0.05;~0.08$ and  
$\rho_{r} \cong 9.7;~9.4;~8.5~{\rm g/cm^3}$, respectively.
As can be easily seen from Figs. 1a - 1c, Fig. 2a and Fig. 3a,
all the three points lie rather close to, but outside the regions 
determined by the inequality (10b) (or (18c)) and by the inequality (14). 
Thus, for $\sin^22\theta < 0.040$ 
conditions (9) are never exactly satisfied for the 
trajectories of neutrinos crossing the Earth core \cite{SP3198}:
one has either $\Delta E'X' = \pi$ and $\Delta E''X''$ relatively close but
different from $\pi$, or {\it vice versa}. 
This can also be seen from the fact that at $\sin^22\theta \ltap 0.02$,
one of the two conditions in (9) is equivalent to the following
physical condition \cite{SP3198}:
$$ \pi~\left [ {1\over {X'}} + {1\over {X''}} \right ] 
\cong \sqrt{2} G_F {1\over{m_N}}(Y_e^{c}\bar{\rho}_{c} - 
Y_e^{man}\bar{\rho}_{man}) = 
\sqrt{2} G_F (\bar{N}_e^{c} - \bar{N}_e^{man}),~~~\eqno(22)$$ 
  
\noindent $\bar{N}_e^{c}$ and  $\bar{N}_e^{man}$ being the average
core and mantle electron number densities along 
the chosen neutrino trajectory. 
One can easily convince oneself using eq. (8), the values of
the core and Earth radii $R_{c}$ and $R_{\oplus}$ and the values of
$Y_e^{c}$, $\bar{\rho}_{c}$, $Y_e^{man}$ and $\bar{\rho}_{man}$ that the 
above equality cannot be exactly satisfied for the 
trajectories of neutrinos crossing the Earth core.

   Conditions (9) are fulfilled
for certain values of $\sin^22\theta$ from the interval 
$(0.04 - 0.10)$, but in this case inequality (10b) 
(or (18c)) and inequality (14), 
which guarantee that the relevant neutrino transition probabilities have
maxima, are not fulfilled. Nevertheless, the probabilities
$P_{e2}$ and $P(\nu_{\mu} \rightarrow \nu_{e})$ 
are strongly (resonantly) enhanced even when conditions (9) 
are only approximately satisfied due to the fact that the
neutrino oscillation length resonance is relatively wide \cite{SP3198}.

  As Figs. 1a - 1c, 2a and 3a demonstrate,
the point where $\Delta E'X' = \pi$ and $\Delta E''X'' = \pi$,
which is closest to the regions determined by (10a), (10b) or by (14),
corresponds to $h = 0^{\circ}$. As a consequence, the values of
$P_{e2}$ and $P(\nu_{\mu} \rightarrow \nu_{e})$ 
at their corresponding
maxima for the trajectory with $h = 0^{\circ}$ and small mixing angles
($\sin^22\theta \ltap 0.02$)
are bigger than $max~P_{e2}$ and $max~P(\nu_{\mu} \rightarrow \nu_{e})$  
for the other trajectories crossing the Earth core: 
$max~P_{e2}$ and $max~P(\nu_{\mu} \rightarrow \nu_{e})$ decrease
with the increase of $h$. This is illustrated in Figs. 4a and 4b showing
the dependence of the probability $P_{e2}$ on $h$ for 
$\rho_{r} = 10~{\rm g/cm^3}$ (Fig. 4a) and on $h$ and $\rho_r$ (Fig. 4b)
for $\sin^22\theta = 0.01$. Figure 4a clearly demonstrates also that
i) the neutrino oscillation length resonance can take place
only if the neutrinos cross the Earth core, 
ii) this resonance is relatively wide in the Nadir angle 
variable $h$ as well, and that iii) 
the resonance of interest is distinctly different from the MSW
resonance in the mixing as the enhancement of $P_{e2}$ due to the 
latter cannot and does not 
take place at $\rho_{r} = 10~{\rm g/cm^3}$ \cite{SP3198}.  

   Comparing Fig. 2a (2b) and Fig. 3a (3b) one notices 
that the regions allowed by the
inequalities (10b) or (18c) and (14) for values of the resonance densities
from the interval (19a) differ somewhat even for 
$\sin^22\theta \cong (0.01 - 0.10)$. This difference is due to the presence 
of $\theta$ in the argument of the cosine in (18c). As a result,
the points at which we have $\Delta E'X' = \pi$ and $\Delta E''X'' = \pi$
for the different trajectories through the Earth core are closer to the
region where (18c) holds than to the region of validity of (14).
Correspondingly, we can expect that for given $h \ltap 28^{\circ}$,
$\sin^22\theta \cong (0.01 - 0.10)$, and transitions 
involving active neutrinos only, the maxima associated with the neutrino 
oscillation length resonance satisfy
$max~P_{e2} \gtap max~P(\nu_{\mu} \rightarrow \nu_{e})$ and/or that
the width of the resonance in $P_{e2}$ is larger than that in
$P(\nu_{\mu} \rightarrow \nu_{e})$. The dependence of 
the probability $P(\nu_{\mu} \rightarrow \nu_{e})$ on 
$E/\Delta m^2$ and $h$ for three values of $\sin^22\theta$, 0.005, 0.01 and 0.05,
is shown graphically in Figs. 5a - 5c.

   Similar analysis can be performed for the 
$\nu_2 \rightarrow \nu_e$ small mixing angle 
(solar) neutrino transitions in the Earth in the case of
$\nu_e - \nu_s$ mixing. Equations (18a), 
(18c), (19a) - (19b) and
(20) - (22) and the discussions associated 
with them remain valid in this case after a minor change: we have
to replace $Y_e$ by $0.5(3Y_e - 1)$ and use for the ratio $r$ 
the expression
$$r \equiv \frac{\bar{\rho}_{c}~(3Y_e^{c} - 1)}
{\bar{\rho}_{man}(3Y_e^{man} - 1)}~~~\eqno(23)$$
\vskip 0.2truecm
\noindent in these equations. The 
numerical values of the parameter $r$ and 
of the upper and lower limits (18a) 
and in (20) now read, 
respectively: 
$r \cong 2.180$, 
$\rho^{res}_{man} <  15.1~{\rm g/cm^3}$
and $18.3~{\rm g/cm^3} < \rho^{res}_{man} < 30.4~{\rm g/cm^3}$.  
The lines of $\Delta E'X' = \pi (2k +1)$ and 
$\Delta E''X'' = \pi (2k' +1)$ 
in the 
$\Delta m^2/E - \sin^22\theta$ plane 
are shown in Figs. 6a - 6c
for the trajectories with $h = 0^{\circ};~13^{\circ};~23^{\circ}$, 
while the regions where 
conditions (10a) and (10b)
are satisfied 
are depicted in Fig. 7. 
As we have already indicated \cite{SP3198},
the oscillation length resonance conditions (9) and (10a) - (10b) 
are not even approximately fulfilled simultaneously 
for the $\nu_2 \rightarrow \nu_e$ transitions of interest 
at small mixing angles. 
This is well illustrated by Figs. 6a - 6c and 7.
Conditions (9) with $k = k' = 0$ can 
\footnote{Among the lines on the 
$\Delta m^2/E - \sin^22\theta$ plane
determined by the conditions $\Delta E'X' = \pi(2k + 1)$ and 
$\Delta E''X'' = \pi(2k' + 1)$, those corresponding to 
$k = 0$ and $k' = 0$ are located at small mixing angles
closer to the region 
where conditions (10a) and (10b) (or (14)) hold, than the lines for
$k \neq 0$ and $k' \neq 0$.}
both be satisfied only 
for $h \gtap 15^{\circ}$.
At $h = 16^{\circ}$, for example, we have 
$\Delta E'X' = \pi$ and $\Delta E''X'' = \pi$
at $\sin^22\theta \sim 0.01$ for 
$\Delta m^2/E \sim 5.5\times 10^{-7}~{\rm eV^2/MeV}$
($\rho_r \sim 14.3~{\rm g/cm^3}$),
but this point is located rather far from the
region where (10a) - (10b) hold (Fig. 7).
With the increase of $h$ the point where one has 
$\Delta E'X' = \pi$ and 
$\Delta E''X'' = \pi$ is moving further away from the
region where (10a) - (10b) are fulfilled (Fig. 6c).
As a consequence, the effect of the 
oscillation length resonance 
on the $\nu_2 \rightarrow \nu_e$ transitions 
at small $\sin^22\theta$ is considerably weaker
if the latter are triggered by $\nu_e - \nu_s$ 
mixing in vacuum. In spite of that, 
the $\nu_2 \rightarrow \nu_e$
transition probability is noticeably 
enhanced in the region of the MSW effect in 
the Earth core by interference terms present in
$P_{e2}$ in addition to the MSW terms and the dominant term
associated with the oscillation length resonance \cite{SP3198} -
notably by the contribution of the last term in eq. (2).
Analogous results are valid for the $\nu_e \rightarrow \nu_s$
transitions at small mixing.

  It is quite interesting that, 
in contrast to the case just considered, 
the oscillation length resonance 
conditions (9) with $k = 0$ and $k' = 0$ can both be satisfied 
for the $\bar{\nu}_{\mu} \rightarrow \bar{\nu}_s$ 
transitions at small mixing angles 
($\sin^22\theta \ltap 0.10$)
for the trajectories with $h \ltap 15^{\circ}$.
These transitions are strongly enhanced
for $h \ltap 15^{\circ}$
in spite of the fact that the points where the equalities 
$\Delta E'X' = \pi$ and $\Delta E''X'' = \pi$ hold
lie outside the region determined by the inequality (14)
(Figs. 8a, 8b, 9, 10a-10b). 
For $h = 0^{\circ}$, for instance, one has  
$\Delta E'X' = \pi$, $\Delta E''X'' = \pi$
for $\sin^22\theta \cong 0.05$ and
$\Delta m^2/E \cong 6\times 10^{-7}~{\rm eV^2/MeV}$
($\rho_r \cong 15.8~{\rm g/cm^3}$).
However, the dominating absolute maximum of 
$P(\bar{\nu}_{\mu} \rightarrow \bar{\nu}_s)$ 
occurs in the region of the MSW resonance
in the Earth core, i.e., 
at $\Delta m^2/E \sim 4.1\times 10^{-7}~{\rm eV^2/MeV}$
($\rho_r \sim 10.8~{\rm g/cm^3}$).
Actually, at the point where conditions (9) hold
one has $P(\bar{\nu}_{\mu} \rightarrow \bar{\nu}_s) = 
\sin^2(2\theta''_{m} - 4\theta'_{m})$, but the function
$\sin^2(2\theta''_{m} - 4\theta'_{m})$ takes a relatively
small value and $P(\bar{\nu}_{\mu} \rightarrow \bar{\nu}_s)$
is strongly suppressed. At the same time 
in the example we are considering
$max~P(\bar{\nu}_{\mu} \rightarrow \bar{\nu}_s) \cong 0.94$.
The strong enhancement of 
$P(\bar{\nu}_{\mu} \rightarrow \bar{\nu}_s)$ at small mixing angles
is due to the term in the expression for 
$P(\bar{\nu}_{\mu} \rightarrow \bar{\nu}_s)$,
describibg the MSW effect in the Earth core and
to an equally important interference term -
the analog of the last term in eq. (13) for 
$P(\nu_{\mu} \rightarrow \nu_e)$.
Thus, the mechanism of enhancement of 
$P(\bar{\nu}_{\mu} \rightarrow \bar{\nu}_s)$
at small mixing angles is similar to the mechanism
of enhancement of $P_{e2}$ in the case of 
$\nu_e - \nu_s$ mixing.
Note that, e.g., for $h = 0^{\circ}$, $\sin^22\theta \cong 0.01$ and 
$\Delta m^2 = 5\times 10^{-4};~10^{-3};~4\times 10^{-3}~{\rm eV^2}$, 
the dominating absolute maximum
in $P(\bar{\nu}_{\mu} \rightarrow \bar{\nu}_s)$ 
takes place at $E \cong 1.2;~2.3;~9.6~{\rm GeV}$,
which fall in the intervals of energies of 
$\bar{\nu}_{\mu}$ contributing either to the sub-GeV 
or to the multi-GeV $\mu-$like samples of 
Super-Kamiokande atmospheric neutrino events. 
Obviously, the discussed 
strong enhancement of $P(\bar{\nu}_{\mu} \rightarrow \bar{\nu}_s)$
(or of $P(\nu_{\mu} \rightarrow \nu_s)$ if $\Delta m^2 < 0$)
can be utilized to perform a rather 
sensitive search for, and might allow to detect, 
$\bar{\nu}_{\mu} \rightarrow \bar{\nu}_s$ 
(or $\nu_{\mu} \rightarrow \nu_s$)
transitions of, e.g., the atmospheric $\bar{\nu}_{\mu}$ 
(or $\nu_{\mu}$) at small mixing angles. 
For $sin^22\theta = 0.01;~0.05$ the dependence of 
the probability $P(\bar{\nu}_{\mu} \rightarrow \bar{\nu}_s)$
on $h$ and $E/\Delta m^2$ is shown in Figs. 10a and 10b.

    At large mixing angles, $\sin^22\theta \gtap 0.10$,
and $\nu_e - \nu_{\mu (\tau)}$ mixing, inequalities (10a) and (10b)
or (14) can be satisfied only in the interval (19b) (Fig. 2a).
The corresponding regions of maxima of 
$P_{e2}$ and $P(\nu_{\mu} \rightarrow \nu_{e})$ 
differ somewhat 
(compare Fig. 2a (2b) and Fig. 3a (3b)). In the regions 
where (10a) and (10b) hold or
(14) is satisfied, conditions (9) 
are fulfilled only for 
$k = 0$ and $k' = 1$, and only for $h \gtap 20^{\circ}$: 
we have $\Delta E'X' = \pi$ and $\Delta E''X'' = 3\pi$
for the trajectory with $h = 23^{\circ}$, for instance, 
at  $\sin^22\theta \cong 0.88$ and 
$\rho_r \cong 1.23~{\rm g/cm^3}$ ($E/\Delta m^2 \cong 
3.695\times 10^{6}~{\rm MeV/eV^2}$). Correspondingly, 
the probability 
$P(\nu_{\mu} \rightarrow \nu_{e})$
has a maximum at the indicated point due to the neutrino
oscillation length resonance 
\footnote{Conditions (9) (and (14) or (10a) - (10b)) are also 
satisfied for the $\nu_{\mu (e)} \rightarrow \nu_{e(\mu)}$
and $\nu_{2} \rightarrow \nu_{e}$ 
transitions at $\sin^22\theta \cong 1$. The study of this 
specific case will be reported elsewhere.}.

   Conditions (9) can be satisfied for        
$k \geq 0$ and $k' \geq 1$ at large mixing angles 
in the region $\rho^{res}_m > \bar{\rho}_m$ as well,
as Figs. 1a - 1c illustrate. For the trajectory with 
$h = 13^{\circ}$, for example, we have 
$\Delta E'X' = \pi$ and $\Delta E''X'' = 3\pi$
at  $\sin^22\theta \cong 0.52$ and
$\rho_r \cong  5.4~{\rm g/cm^3}$;
$\Delta E'X' = 3\pi$ and $\Delta E''X'' = 5\pi$
at  $\sin^22\theta \cong 0.4$ and
$\rho_r \cong 15~{\rm g/cm^3}$; and 
$\Delta E'X' = 5\pi$ and $\Delta E''X'' = 9\pi$
at  $\sin^22\theta \cong 0.42$ and
$\rho_r \cong 23~{\rm g/cm^3}$. All these points
correspond, however, to minima in 
$P_{e2}$: now conditions (12) (instead of (10a)) and 
(10b) are fulfilled.
They correspond to saddle points in
$P(\nu_{\mu} \rightarrow \nu_{e})$.
Some of the minima can be very deep.
This is clearly seen in Fig. 11 showing
the dependence of $P_{e2}$ on 
$\rho_r$ for  $\sin^22\theta = 0.4$. 
Only the minimum at $\rho_r \sim (6.0 - 7.0)~{\rm g/cm^3}$
($h = 0^{\circ};~13^{\circ}$) is an absolute one: we have $P_{e2} = 0$ 
at this minimum. Note that such a deep minimum is not 
present in the same region for the trajectory with $h = 23^{\circ}$, 
in accord with the results shown in Figs. 1a - 1c.
Nevertheless, the indicated minimum in $P_{e2}$ at $h \ltap 16^{\circ}$  
leads to the ``island'' of somewhat reduced values of
the D-N asymmetry in the vicinity of the point
$\sin^22\theta \cong 0.45$,
$\Delta m^2 \cong 5\times 10^{-5}~{\rm eV^2}$  
in the {\it core} sample of the solar
neutrino events in the Super-Kamiokande detector
(see Fig. 3c in \cite{Art2}). The wide deep minimum in
$P_{e2}$ in the case of $\nu_{e} \rightarrow \nu_{s}$ transitions 
of the core-crossing solar neutrinos, taking place  
at $\sin^22\theta \cong (0.30 - 0.55)$ and
$\rho_r \cong  (6 - 10)~{\rm g/cm^3}$
(Fig. 12), has apparently a similar origin   
\footnote{The existence of this minimum is reflected 
in the prediction of a spectacularly 
large in absolute value but {\it negative} D-N asymmetry
at $\sin^22\theta \sim 0.50$,
$\Delta m^2 \sim 3.5\times 10^{-6}~{\rm eV^2}$  
in the {\it core} sample of the
the Super-Kamiokande
solar neutrino induced events (see footnote 11 and 
Figs. 3.11, 3.12 and 6a - 6d 
in \cite{Art3}).}.  
The other minima in $P_{e2}$ at  
$\rho_r > 10~{\rm g/cm^3}$ are relatively shallow.   

  It should be noted that the large mixing angle (LMA)
MSW $\nu_{e} \rightarrow \nu_{\mu (\tau)}$ transition 
solution of the solar neutrino problem is possible,
as the analyzes of the solar neutrino data show,
for $\Delta m^2 \gtap 10^{-5}~{\rm eV^2}$ and 
$0.60 \ltap \sin^22\theta \ltap 0.96$.
As we have seen, the oscillation length resonance maxima
in $P_{e2}$ can take place at large mixing angles only for 
$\rho_r < \bar{\rho}_{man} \cong 4.5~{\rm g/cm^3}$.
Correspondingly, $P_{e2}$ can be enhanced due to this resonance
only for neutrino energies  
$$E \gtap 30~{\rm MeV}~\cos2\theta,~~~\eqno(24)$$ 

\noindent which can fall in the range of (5 - 14) MeV,
causing substantial D-N asymmetry 
in the spectrum of $^{8}$B neutrinos, only for 
$0.80 \ltap \sin^22\theta \ltap 0.95$ and only for the trajectories
with $h \gtap 20^{\circ}$. 
Figure 13 illustrates the 
dependence of the probability $P_{e2}$
on $\rho_{r}$ for $\sin^22\theta = 0.9$.

  Let us emphasize in  conclusion of this 
Section that for most of the neutrino trajectories 
crossing the Earth core the neutrino oscillation 
length resonance takes place at values of the resonance
density which differ from the values of the density
met, according to the Earth models, 
in the Earth mantle and in the Earth core.

\vglue 0.3cm
\leftline{\bf 3.2. Properties of the Probabilities $P_{e2}$, 
$P(\nu_{\mu} \rightarrow \nu_{e})$ 
($P(\nu_{e} \rightarrow \nu_{\mu (\tau)})$) 
and $P(\bar{\nu}_{\mu} \rightarrow \bar{\nu}_{s})$}

\hskip 0.1truecm {\bf in the Oscillation Length Resonance Regions}
\vglue 0.2cm

  As one can easily check utilizing Figs. 1d - 1f, 2b and 3b, 
practically all maxima of  $P_{e2}$ and $P(\nu_{\mu} \rightarrow \nu_{e})$
associated with the neutrino oscillation length 
resonance, take place along the line 
$\Delta E'X' = \pi$. For $\Delta E'X' = \pi(2k + 1)$,
the probabilities  $P_{e2}$ and $P(\nu_{\mu} \rightarrow \nu_{e})$
have for any $\theta$ the following form \cite{SP3198}:
$$P_{e2} = \sin^2\theta~ +~ {1\over {2}} 
\left [1 - \cos \Delta E''X'' \right ] \left [ \sin^2(2\theta''_{m} - 
4\theta'_{m} + \theta) - \sin^2\theta \right ],
~\Delta E'X' = \pi(2k + 1),~~\eqno(25)$$
$$P(\nu_{\mu} \rightarrow \nu_{e}) = 
P(\nu_{e} \rightarrow \nu_{\mu (\tau)}) = {1\over {2}} 
\left [1 - \cos \Delta E''X'' \right ]\sin^2(2\theta''_{m} - 
4\theta'_{m}),~~\Delta E'X' = \pi(2k + 1).~\eqno(26)$$

\noindent Expressions (25) and (26) describe 
$P_{e2}$ and $P(\nu_{\mu} \rightarrow \nu_{e})$ in the region of 
the neutrino oscillation length resonance; they determine, in particular,
the width of the resonance. Obviously, many of the properties of, e.g.,
the probability $P_{e2}$ in the resonance region are determined by the 
properties of the function 
$\sin^2(2\theta''_{m} - 4\theta' + \theta)$ in eq. (25). 
 
 The dependence 
of the function $\sin^2(2\theta''_{m} - 
4\theta'_{m} + \theta)$ on $\rho_r$ 
is shown graphically in Fig. 14 
for $\sin^22\theta = 0.005;~0.01;~0.05;~0.10;~0.50;~0.90$.
At small mixing angles, $\sin^22\theta \ltap 0.02$,
$\sin^2(2\theta''_{m} - 4\theta'_{m} + \theta)$
has three distinct maxima \cite{SP3198}: at 
$\rho^{res}_{c} \cong \bar{\rho}_{c}$
where $2\theta''_m \cong \pi/2$ and
$4\theta'_m - \theta \cong  
5.6\theta \ltap 0.13\pi \ll 2\theta''_m$, and 
at the points where 
$$2\theta'_m \cong \frac{\pi}{4},~~\frac{3\pi}{4},~~\eqno(27a)$$ 

\noindent which is realized for
$$\rho^{res}_{man} \cong \frac{\bar{\rho}_{man}}
{1 \pm 2\sqrt{2} \theta}~~.~~~~\eqno(27b)$$
\vskip 0.3truecm
\noindent At these two points $2\theta''_m \cong \pi$.
The same function has two distinct minima in the interval
$0 < \rho^{res}_{man} < 
\bar{\rho}_{c}$: at $\rho^{res}_{man} \cong \bar{\rho}_{man}$
($4\theta'_m \cong \pi$, $2\theta''_m \cong \pi$) and at \cite{SP3198}
$$\rho^{res}_{man} \cong \kappa \bar{\rho}_{man},~~~\eqno(28a)$$ 

\noindent where
$$\kappa \cong \frac{r\sqrt{2} + \sqrt{r}} {\sqrt{2} + \sqrt{r}}~.~~~ 
~~\eqno(28b)$$

\vskip 0.3truecm
\noindent The first minimum is rather deep, while the second 
one is relatively shallow \cite{SP3198}.
At the first minimum
$\sin^2(2\theta''_{m} - 4\theta'_{m} + 
\theta) \cong \sin^2\theta$, and at the second
$$min~\sin^2(2\theta''_{m} - 4\theta'_{m} + \theta) \cong 
\sin^2\left [ \pi - 2\theta \kappa
((r -  \kappa)^{-1} + 2(\kappa - 1)^{-1}) + 
 \theta \right ].~~\eqno(28c)$$

\noindent For $r \cong 2.44$ ($\nu_e - \nu_{\mu (\tau)}$ mixing) 
one finds $\kappa \cong 1.68$, 
and correspondingly $\rho^{res}_{man} = 
\kappa \bar{\rho}_{man} \cong 7.57~{\rm g/cm^3}$ 
and $min~\sin^2(2\theta''_{m} - 4\theta'_{m} + \theta) \cong 
\sin^2(\pi - 13.3\theta)$, which is in 
excellent agreement with our numerical results
\footnote{If $\nu_e - \nu_{s}$ mixing takes place
we have $r \cong 2.18$, $\kappa \cong 1.58$ and 
this minimum occurs at $\rho^{res}_{man} \cong 7.1~{\rm g/cm^3}$.
In the case of $\bar{\nu}_{\mu} \rightarrow \bar{\nu}_{s}$ transitions,
$r \cong 2.67$ and $\kappa \cong 1.76$, so 
the minimum is at $\rho^{res}_{man} \cong 8.0~{\rm g/cm^3}$.}.
Note that the positions of both minima depend
very weakly on $\theta$ at small mixing angles. 

  The oscillation length resonance maximum in $P_{e2}$ occurs
for given $\sin^22\theta \ltap 0.05$ and the different neutrino
trajectories through the Earth core at values of 
$\rho_r \cong (8.0 - 9.7)~{\rm g/cm^3}$. The precise position of the
maximum is determined by the factor
$0.5(1 - \cos \Delta E''X'')$ in eq. (25) and it does not coincide 
with the position of the maximum of the function
$\sin^2(2\theta''_{m} - 4\theta'_{m} + \theta)$ \cite{SP3198}.
Actually, the maximum of $P_{e2}$ falls 
in the interval of values of $\rho_r$
between the minimum at $\rho^{res}_{man} \cong \kappa \bar{\rho}_{man}
\cong 7.57~{\rm g/cm^3}$ and the maximum at 
$\rho^{res}_{c} \cong \bar{\rho}_{c} \cong 11.5~{\rm g/cm^3}$ of
the factor $[\sin^2(2\theta''_{m} - 
4\theta'_{m} + \theta) - \sin^2\theta]$, where the latter
rises steeply (Fig. 14). For this reason, in particular,
the width of the neutrino oscillation length
resonance in $P_{e2}$ is determined at $\sin^22\theta \ltap 0.05$
by the expression (25) and not only by the function
$\sin^2(2\theta''_{m} - 4\theta'_{m} + \theta)$, as one 
might have expected on the basis of eq. (11).
The same considerations apply to the position of the 
oscillation length resonance maximum and the width of the resonance
in the probability $P(\nu_{\mu} \rightarrow \nu_{e})$ 
in the indicated range of values of $\sin^22\theta$.

   The minimum of $\sin^2(2\theta''_{m} - 4\theta'_{m} + \theta)$
at $\rho^{res}_{man} \cong \kappa \bar{\rho}_{man}$
becomes shallower as $\sin^22\theta$ increases 
starting, say, from the value 0.001.
At $\sin^22\theta = 0.05$ ($\theta \cong 0.11$)
it is still visible (Fig. 14), but 
$min~\sin^2(2\theta''_{m} - 4\theta'_{m} + \theta) \cong 
0.98$. As $\sin^22\theta$ increases further
this minimum disappears and the two maxima it separated, at
$\rho^{res}_{man} \cong \bar{\rho}_{man}/
(1 - 2\sqrt{2} \theta)$ ($2\theta_{m}' \cong \pi/4$) and at 
$\rho^{res}_{c} \cong \bar{\rho}_{c}$
($2\theta_{m}'' \cong \pi/2$),
merge to form one maximum. The latter is located practically at the position 
of the indicated minimum, i.e., at 
$\rho^{res}_{man} \cong \kappa \bar{\rho}_{man}$,
as this is clearly seen in the case of   
$\sin^22\theta = 0.10$ in Fig. 14: at this point 
$2\theta''_{m} - 4\theta'_{m} + \theta$ is just close to $\pi/2$ due to the 
relatively large value of $\theta$. As 
$\sin^22\theta$ increases beyond 0.10, this maximum diminishes
since $2\theta''_{m} - 4\theta'_{m} + \theta$ decreases
and at $\sin^22\theta = 0.50$, where 
$2\theta''_{m} - 4\theta'_{m} + \theta \cong 0$, it does not exist.

  The maximum of $\sin^2(2\theta''_{m} - 4\theta'_{m} + \theta)$ 
in the region $\rho^{res}_{man} \sim \kappa \bar{\rho}_{man} 
\cong 7.57~{\rm g/cm^3}$ for $\sin^22\theta \sim (0.05 - 0.10)$
is remarkably wide (Fig. 14). However, the oscillating factor
$0.5(1 - \cos \Delta E''X'')$  in eqs. (25) and (26)
goes through zero in the same region, 
thus reducing considerably the effect of the large width 
of this maximum on the probabilities $P_{e2}$,
$P(\nu_{\mu} \rightarrow \nu_{e})$ and 
$P(\bar{\nu}_{\mu} \rightarrow \bar{\nu}_{s})$ (Figs. 5c and 10b).

The position of the minimum of 
$\sin^2(2\theta''_{m} - 4\theta'_{m} + \theta)$  
at $\rho^{res}_{man} \cong \bar{\rho}_{man}$, where
$4\theta'_{m} \cong \pi$, is very stable with respect to the 
change of $\sin^22\theta$ up to 
$\sin^22\theta \cong 0.5$.

  The maximum and the minimum of 
$\sin^2(2\theta''_{m} - 4\theta'_{m} + \theta)$, which
at small mixing angles are located at 
$\rho^{res}_{man} \cong \bar{\rho}_{man}/
(1 + 2\sqrt{2} \theta)$ and at 
$\rho^{res}_{man} \cong \bar{\rho}_{man}$, respectively,
``survive'' at large mixing angles, but change
their positions moving to somewhat smaller values of 
$\rho^{res}_{man}$. These are the only extrema 
$\sin^2(2\theta''_{m} - 4\theta'_{m} + \theta)$ has
at large mixing angles. At $\sin^22\theta \gtap 0.4$ 
the function $[\sin^2(2\theta''_{m} - 
4\theta'_{m} + \theta) - \sin^2\theta]$
``shrinks'' essentially to the region
$\rho^{res}_{man} \ltap \bar{\rho}_{man}$,
where it can take relatively large values (Fig. 14).
This is also the region where the two extrema of 
$[\sin^2(2\theta''_{m} - 
4\theta'_{m} + \theta) - \sin^2\theta]$ are located: the maximum
occurs at $\rho_r \ltap 2.0~{\rm g/cm^3}$, while the minimum takes 
place at $\rho_r \cong (3.5 - 4.5)~{\rm g/cm^3}$, depending on the
neutrino trajectory. Outside the indicated region
$[\sin^2(2\theta''_{m} - 
4\theta'_{m} + \theta) - \sin^2\theta]$ is close to 0.
 
    At $\rho^{res}_{c} \gg \bar{\rho}_{c}$ and at 
$\rho^{res}_{man} \ll \bar{\rho}_{man}$ 
we have for small mixing angles 
$\sin^2(2\theta''_{m} - 4\theta'_{m} + \theta) \cong 
\sin^2(\theta (1 - \epsilon))$, where $\epsilon > 0$
is a small quantity, $\epsilon \ll 1$, so that 
$\sin^2(\theta (1 - \epsilon)) \cong \sin^2 \theta $.
Nevertheless, one always has
$\sin^2(2\theta''_{m} - 4\theta'_{m} + \theta) < 
\sin^2 \theta$  in the indicated regions, 
although the difference between the
two sines is very small. Obviously, the probabilities
$P_{e2}$ and $P(\nu_{\mu} \rightarrow \nu_{e})$ are 
rather strongly suppressed 
at small mixing angles
when $\rho^{res}_{c} \gg \bar{\rho}_{c}$ or 
$\rho^{res}_{man} \ll \bar{\rho}_{man}$.  
The same conclusion is valid at large mixing angles, 
$\sin^22\theta \gtap 0.2$, for the probability $P_{e2}$;
it is valid also in the region $\rho^{res}_{man} \ll \bar{\rho}_{man}$
for the probability $P(\nu_{\mu} \rightarrow \nu_{e})$.
However, as can be shown (and Fig. 14 suggests), 
$P(\nu_{\mu} \rightarrow \nu_{e})$
can take relatively large values at $\sin^22\theta \gtap 0.2$
for $\rho^{res}_{c} \gg \bar{\rho}_{c}$. The 
behavior of the probability 
$P(\nu_{\mu} \rightarrow \nu_{e})$
as a function of $\rho_r$ for several different values of 
$\sin^22\theta$ and for $h = 0^{\circ};~13^{\circ};~23^{\circ}$ is illustrated
in Figs. 15a - 15c. 

  Let us note finally that the probabilities $P_{e2}$ in the case of 
$\nu_e - \nu_s$ mixing, $P_{e2,s}$, $P(\nu_{e} \rightarrow \nu_{s})$
and $P(\bar{\nu}_{\mu} \rightarrow \bar{\nu}_{s})$ are 
given by somewhat more complicated expressions 
than (25) and (26) in the regions of their strong (resonance-like) 
enhancement at small mixing angles \cite{SP3198}:
$$P^{res}_{e2,s} \cong
{1\over {2}}
\left [1 - \cos \Delta E''X'' \right ] \sin^2 (2\theta''_{m} - \theta)
~~~~~~~~~~~~~~~~~~~~~~~~~~~~~~~~~~~~~~~~~~~~~~~~~~~~~~~~~~~~~~~~~~~~~~~~~~~~~~~$$
$$~~~~~~+~ {1\over {4}} \left [1 - \cos \Delta E''X'' \right ]
\left [1 - \cos \Delta E'X' \right ] 
\left [ \sin^2(2\theta''_{m} - 4\theta'_{m}  + \theta) - 
 \sin^2(2\theta''_{m} - \theta) \right ]~~~~~~~~~~~~~~~~~~~$$
$$~~~~+~{1\over {4}} \left [ \cos (\Delta E'X' - \Delta E''X'')
- \cos (\Delta E'X' + \Delta E''X'') \right ]
 \sin (4\theta'_{m} - 2\theta)
 \sin (2\theta''_{m} - 2\theta'_m).\eqno(29)$$
\noindent The expressions for 
 $P(\nu_{e} \rightarrow \nu_{s})$ and 
 $P(\bar{\nu}_{\mu} \rightarrow \bar{\nu}_{s})$ can be obtained
 by setting formally $\theta = 0$, but keeping 
 $\theta'_{m} \neq 0$ and $\theta''_{m} \neq 0$
 in eq. (29) and using the definitions  
 of $\rho^{res}_{man,c}$ given at the end of Section 2.
The first and the last terms in eq. (29) 
and the corresponding terms in the expressions for 
$P(\nu_{e} \rightarrow \nu_{s})$
and $P(\bar{\nu}_{\mu} \rightarrow \bar{\nu}_{s})$
give the dominant contributions. Obviously, they determine the properties
of $P_{e2,s}$, $P(\nu_{e} \rightarrow \nu_{s})$
and $P(\bar{\nu}_{\mu} \rightarrow \bar{\nu}_{s})$ in the
region of interest.   
 
\vglue 0.2cm
\leftline{\bf 4. Conclusions}
\vskip 0.2cm 
 
  The neutrino oscillation length resonance should be
present in the $\nu_{2} \rightarrow \nu_{e}$
transitions taking place when the solar neutrinos
cross the Earth core on the way to the detector,
if the solar neutrino problem is due to 
small mixing angle MSW 
$\nu_{e} \rightarrow \nu_{\mu}$ 
transitions in the Sun.     
The same resonance should be operative also
in the $\nu_{\mu} \rightarrow \nu_{e}$
($\nu_{e} \rightarrow \nu_{\mu (\tau)}$) 
small mixing angle transitions 
of the atmospheric neutrinos crossing the Earth
core if the atmospheric $\nu_{\mu}$ and 
$\bar{\nu}_{\mu}$ indeed take part 
in large mixing vacuum $\nu_{\mu} \leftrightarrow \nu_{\tau}$,
$\bar{\nu}_{\mu} \leftrightarrow \bar{\nu}_{\tau}$ 
oscillations with 
$\Delta m^2 \sim (5\times 10^{-4} - 5\times 10^{-3})~{\rm eV^2}$,
as is strongly suggested by the Super-Kamiokande
atmospheric neutrino data, and if all 
three flavour neutrinos are mixed in vacuum.
The existence of three-flavour-neutrino mixing  
in vacuum is a very natural possibility in view of the 
present experimental evidences for oscillations/transitions of
the flavour neutrinos. In both cases the oscillation 
length resonance produces  
a strong enhancement of the 
corresponding transitions probabilities,
making the transitions observable 
even at rather small mixing angles \cite{SP3198}.
Actually, the resonance may have already
manifested itself in the excess of e-like events 
at $-1 \leq \cos\theta_{z}\leq -0.6$ observed 
in the Super-Kamiokande 
atmospheric 
neutrino data \cite{SP3198,SPNE98,SPal98}.
And it can be responsible for at least part of the
strong zenith angle dependence present in the
Super-Kamiokande multi-GeV and sub-GeV $\mu-$like data 
\cite{SP3198,SPNE98,SPal98}.

  In view of the important role the 
oscillation length resonance can 
play in the interpretation of the results of the experiments
with solar and atmospheric neutrinos 
as well as in obtaining information about 
possible small mixing angle oscillations/transitions of the
(atmospheric) $\nu_{\mu}$ ($\bar{\nu}_{\mu}$) and 
$\nu_{e}$ ($\bar{\nu}_{e}$),
we have performed in the present article   
a rather detailed study of the conditions under which
the resonance 
takes placs in the $\nu_{2} \rightarrow \nu_{e}$, and in the
$\nu_{\mu} \rightarrow \nu_{e}$
($\nu_{e} \rightarrow \nu_{\mu (\tau)}$) and 
$\bar{\nu}_{\mu} \rightarrow \bar{\nu}_{s}$
transitions of neutrinos in the Earth. 
We have examined also some of 
the properties of the resonance.
This was done 
under the simplest assumption
of existence of two-neutrino mixing
(with nonzero-mass neutrinos) in vacuum:
$\nu_e - \nu_{\mu (\tau)}$ or 
$\nu_e - \nu_{s}$, or else
$\nu_{\mu} - \nu_{s}$.
However, in many cases of practical
interest the probabilities of the 
relevant three- (or four-) neutrino mixing
transitions in the Earth reduce effectively to two-neutrino
transition probabilities for which our results are valid.
Thus, our findings have a wider application than it may seem.

   Using our Figs. 1a - 1f, 2a - 3b, 6a - 8c and 9, in particular, 
one can determine practically all relevant values of the 
neutrino oscillation parameters, $E$, $\Delta m^2$ and
$\sin^22\theta$, at which the oscillation length resonance 
occurs for a given neutrino trajectory through the Earth core
in the cases of $\nu_{2} \rightarrow \nu_{e}$, 
$\nu_{\mu} \rightarrow \nu_{e}$
($\nu_{e} \rightarrow \nu_{\mu (\tau)}$) and 
$\bar{\nu}_{\mu} \rightarrow \bar{\nu}_{s}$ transitions.
The sets of such values for the 
transitions of solar and atmospheric neutrinos can be
easily identified. We have shown, for instance,
that the oscillation length resonance conditions (9) 
with $k = 0$ and $k' = 0$ 
can both be satisfied at small $\sin^22\theta$ 
for the $\nu_{2} \rightarrow \nu_{e}$ 
transitions in the case of $\nu_{e} - \nu_{s}$ mixing
only for $h \gtap 15^{\circ}$ (Figs. 6a - 6c). 
In contrast, conditions (9) can be simultaneously fulfilled
at $\sin^22\theta \ltap 0.10$ for the 
$\bar{\nu}_{\mu} \rightarrow \bar{\nu}_s$ 
transitions at $h \ltap 15^{\circ}$ (Figs. 8a - 8c).
However, the points where conditions (9) hold 
lie outside the regions determined 
by the inequalities (10a) - (10b) and by (14).
Nevertheless, as like $P_{e2,s}$ \cite{SP3198}, the probability of
the $\bar{\nu}_{\mu} \rightarrow \bar{\nu}_s$ 
transitions is strongly enhanced at small mixing angles
in the region of the MSW resonance
in the Earth core (Figs. 10a - 10b).
This resonance-like enhancement is due to the interplay of certain
interference terms present in the expression for
$P(\bar{\nu}_{\mu} \rightarrow \bar{\nu}_s)$ - 
notably by the analog of the last term in eq. (13), 
and the term describing the MSW effect in the Earth core.
For, e.g., $h = 0^{\circ}$, $\sin^22\theta \cong 0.01$ and
$\Delta m^2 = 5\times 10^{-4};~10^{-3};~4\times 10^{-3}~{\rm eV^2}$, 
the dominating absolute maximum
in $P(\bar{\nu}_{\mu} \rightarrow \bar{\nu}_s)$ 
takes place at $E \cong 1.2;~2.3;~9.6~{\rm GeV}$,
which fall in the intervals of energies of 
$\bar{\nu}_{\mu}$ contributing either to the sub-GeV 
or to the multi-GeV $\mu-$like samples of 
Super-Kamiokande atmospheric neutrino events. 
The strong (resonance-like) enhancement of  
$P(\bar{\nu}_{\mu} \rightarrow \bar{\nu}_s)$
(or of $P(\nu_{\mu} \rightarrow \nu_s)$ if $\Delta m^2 < 0$)
found in this study can make possible the detection of the
$\bar{\nu}_{\mu} \rightarrow \bar{\nu}_s$ 
(or $\nu_{\mu} \rightarrow \nu_s$)
transitions of, e.g., the atmospheric $\bar{\nu}_{\mu}$ 
(or $\nu_{\mu}$) at small mixing angles.

  We have also investigated the behavior of the probabilities 
$P_{e2}$ and $P(\nu_{\mu} \rightarrow \nu_{e})$ 
($P(\nu_{e} \rightarrow \nu_{\mu (\tau)}$) 
in the region of the neutrino oscillation length resonance.
Simple analytic expressions which 
describe $P_{e2}$ and $P(\nu_{\mu} \rightarrow \nu_{e})$ 
($P(\nu_{e} \rightarrow \nu_{\mu (\tau)}$) 
in the region of the resonance
and determine, in particular, the width of the resonance, 
and the probabilities $P_{e2,s}$, $P(\nu_{e} \rightarrow \nu_{s}$ and 
$P(\bar{\nu}_{\mu} \rightarrow \bar{\nu}_s)$ 
in the region of their resonance-like enhancement 
at small mixing angles, are 
identified. We have shown 
that at small mixing angles 
i) the resonance is relatively wide not only
in the variable $E/\Delta m^2$, but also in the
Nadir angle variable $h$, and that ii)
the values of $P_{e2}$ and $P(\nu_{\mu} \rightarrow \nu_{e})$ 
($P(\nu_{e} \rightarrow \nu_{\mu (\tau)}$) 
at the maxima associated with the 
resonance decrease monotonically with the
increase of $h$. Thus, at small mixing angles 
the enhancement of these probabilities due to the
neutrino oscillation length resonance 
is maximal for the Earth center crossing trajectories,
i.e., for $h = 0^{\circ}$. These conclusions are valid for the
enhancement of the probabilities 
$P_{e2,s}$, $P(\nu_{e} \rightarrow \nu_{s})$ and 
$P(\bar{\nu}_{\mu} \rightarrow \bar{\nu}_s)$ as well.

   As the results obtained in refs. \cite{SP3198,Art2} 
and the present study indicate, it is not excluded that some of 
the current or future high statistics solar and/or atmospheric 
neutrino experiments will be able to 
observe directly the neutrino oscillation length resonance. 
Experiments located at lower geographical latitudes 
than the existing ones or those under construction 
\cite{Rosen97} would be better suited
for this purpose. An atmospheric neutrino detector
capable of reconstructing with relatively good 
precision, e.g.,, the energy and 
the direction of the momentum of the 
incident neutrino in each neutrino-induced event,
would allow to study in detail the 
effects of the oscillation length resonance
in the transitions/oscillations of 
the atmospheric neutrinos crossing the Earth.

\vglue 0.4cm
\leftline{\bf Acknowledgements.}
Part of the work of S.P.T. for the present study was done
at the Aspen Center for Physics during the Workshop on
Neutrino Physics and Astrophysics (June 15 - July 12, 1998).
S.T.P. would like to thank the organizers of the Workshop, and especially
L. Wolfenstein, for the kind hospitality extended to him, and 
L. Wolfenstein, B. Kayser, H. Goldberg, T. Weiler, G. Fuller, A. Balantekin, 
J. Beacom and F. Vissani for the interest 
in the present study and useful discussions. 
This work was supported in part by the Italian MURST
under the program ``Fisica Teorica delle 
Interazioni Fondamentali''  and by Grant PH-510 from the
Bulgarian Science Foundation.


\newpage
\bec{\Large\bf{Figure Captions}}\eec
\noindent
{\bf Figures 1a - 1f.} Lines in the 
$\sin^22\theta - \rho_r$ (a - c)  and
$\sin^22\theta - \Delta m^2/E$ (d - f) planes on which conditions
$\Delta E'X' = \pi (2k + 1),~k=0,1,2,...$ (dotted lines), and 
$\Delta E''X'' = \pi (2k' + 1),~k'=0,1,2,...$ (black solid lines), are
fulfilled for $h = 0^{\circ}$ (center crossing, (a),(d)), $h = 13^{\circ}$ 
(winter solstice for the Super-Kamiokande detector, (b),(e)), 
and  $h = 23^{\circ}$ ((c),(f)).
The parameters $\rho_r$ and
$\Delta m^2/E$ (the vertical axes in figures (a) - (c) and (d) - (f), 
respectively) are in units
of g/cm$^{3}$ and $10^{-7}~ {\rm eV^2/MeV}$. 
The results shown correspond to 
$\nu_e \rightarrow \nu_{\mu (\tau)}$ and  
$\nu_2 \rightarrow \nu_{e}$ transitions in the case of
$\nu_e - \nu_{\mu (\tau)}$ mixing.

\vspace{0.1cm}
\noindent
{\bf Figures 2a - 2b.} Regions of values of 
$\sin^22\theta$ and $\rho_r$ (a) or
$\Delta m^2/E$ (b) of validity of conditions (10a) 
(A + B), (10b) (A + C), and of both (10a) and (10b) (A) 
for the $\nu_2 \rightarrow \nu_{e}$ transitions
in the case of $\nu_e - \nu_{\mu (\tau)}$ mixing. 
Conditions (10b) and (12) hold simultaneously in the region C. 

\vspace{0.1cm} 
\noindent
{\bf Figures 3a - 3b.} 
Regions of validity of condition (14)
in the $\sin^22\theta - \rho_r$ (a) and 
$\sin^22\theta - \Delta m^2/E$ (b) planes (A) 
for the $\nu_e \rightarrow \nu_{\mu (\tau)}$ transitions.

\vspace{0.1cm}
\noindent
{\bf Figures 4a - 4b.}
The  probability $P_{e2}$ 
as a function of the Nadir angle $h$ ((a), upper frame), 
and of $\rho_r$ and $h$ (b), for $\sin^22\theta = 0.01$
in the case of $\nu_e \rightarrow \nu_{\mu (\tau)}$ transitions
of solar neutrinos. The lower frame in figure (a) shows
the dependence on $h$ of the derivative of 
$P_{e2}$ with respect of $h$ in the Stacey 1977 
model of the Earth.

\vspace{0.1cm}
\noindent
{\bf Figures 5a - 5b.} The probability
$P(\nu_e \rightarrow \nu_{\mu (\tau)}) = 
P(\nu_{\mu} \rightarrow \nu_{e})$
as a function of the Nadir angle $h$  
and $E/\Delta m^2$, for $\sin^22\theta = 0.005$~(a), 0.01 (b) and 0.05 (c).

\vspace{0.1cm}
\noindent
{\bf Figures 6a - 6c.} Lines of validity of the conditions
$\Delta E'X' = \pi (2k + 1),~k=0,1,2,...$ (dotted lines), and 
$\Delta E''X'' = \pi (2k' + 1),~k'=0,1,2,...$ (black solid lines)
in the $\sin^22\theta - \Delta m^2/E$ plane 
in the case of $\nu_{e} \rightarrow \nu_s$ and 
$\nu_{2} \rightarrow \nu_e$ transitions
($\nu_{e} - \nu_{s}$ mixing).
The three figures correspond to
$h = 0^{\circ}$ (a), $h = 13^{\circ}$ (b), 
and $h = 23^{\circ}$ (c). 

\vspace{0.1cm} 
\noindent
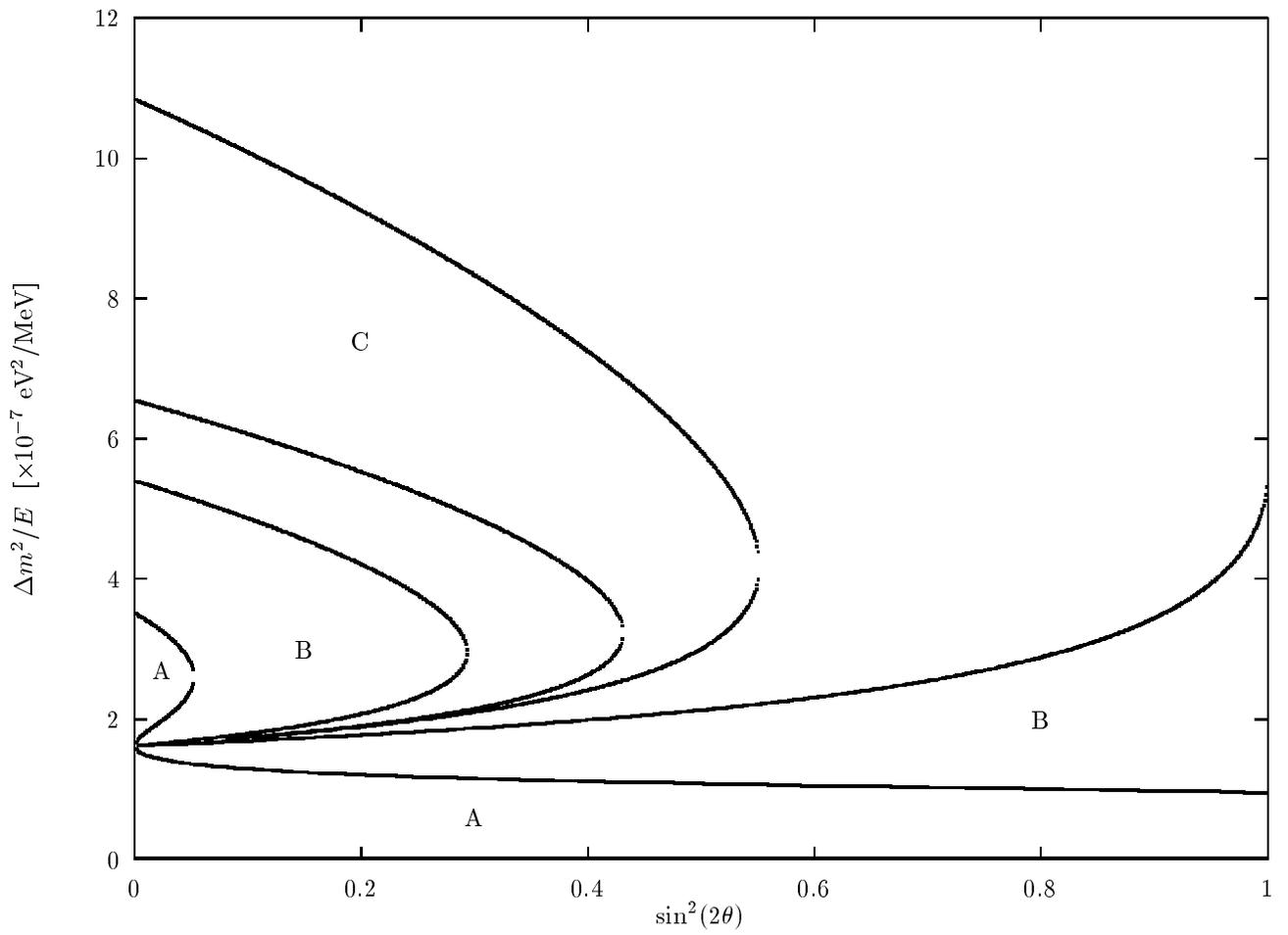
{\bf Figure 7.} 
The same as in figure 2b
for $\nu_{2} \rightarrow \nu_e$ transitions 
induced by $\nu_e - \nu_{s}$ mixing in vacuum.

\vspace{0.1cm}
\noindent
{\bf Figures 8a - 8c.} Lines of validity of the conditions
$\Delta E'X' = \pi (2k + 1),~k=0,1,2,...$ (dotted lines), and 
$\Delta E''X'' = \pi (2k' + 1),~k'=0,1,2,...$ (black solid lines)
in the $\sin^22\theta - \Delta m^2/E$ plane for 
$\bar{\nu}_{\mu} \rightarrow \bar{\nu}_{s}$ transitions and
$h = 0^{\circ}$ (a), $h = 13^{\circ}$ (b), 
and $h = 23^{\circ}$ (c). 

\vspace{0.1cm} 
\noindent
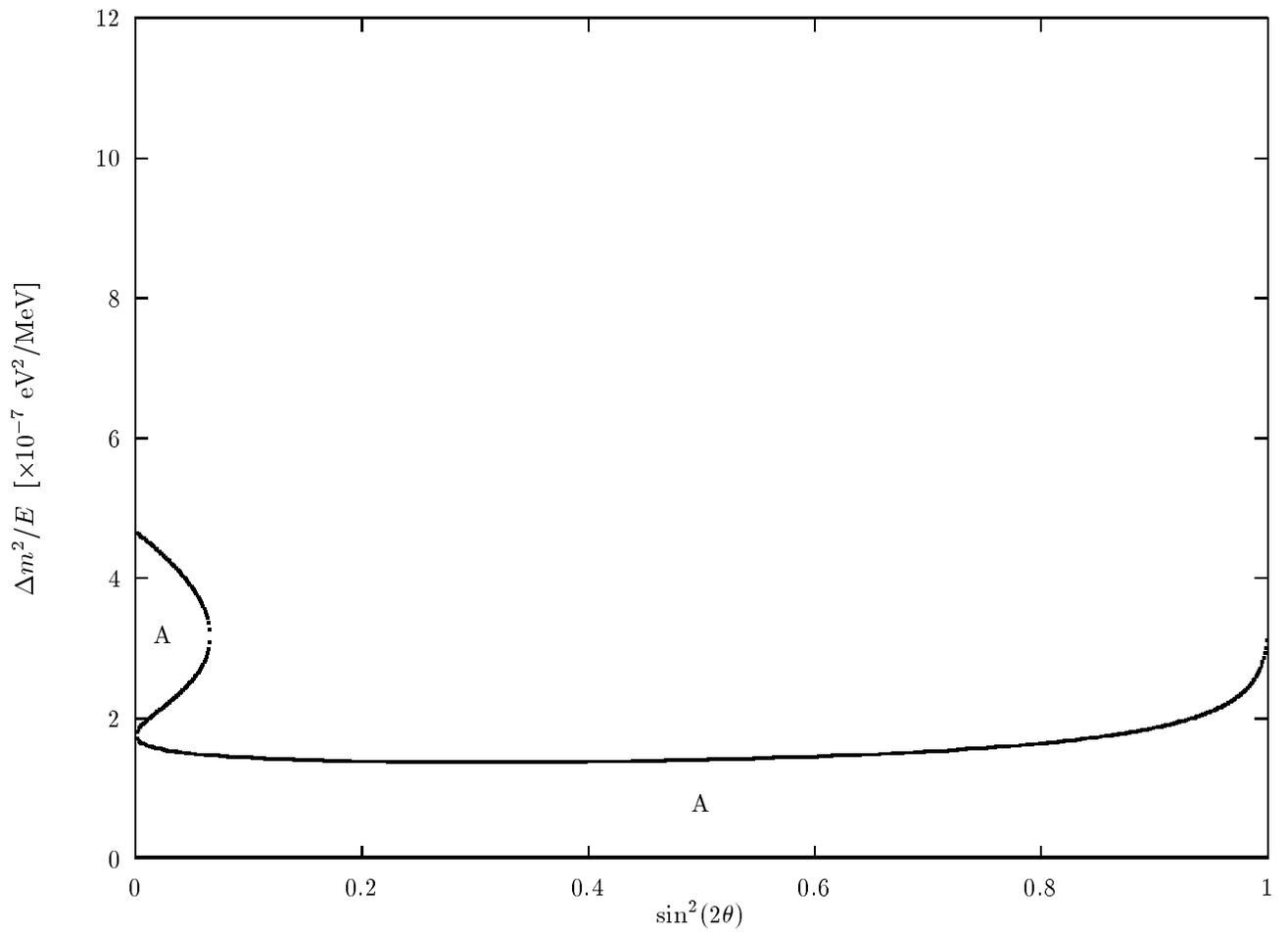
{\bf Figure 9.} 
The same as in figure 3b
for the case of 
$\bar{\nu}_{\mu} \rightarrow \bar{\nu}_{s}$ transitions. 

\vspace{0.1cm}
\noindent
{\bf Figures 10a - 10b.} The probability
$P(\bar{\nu}_{\mu} \rightarrow \bar{\nu}_{s})$
as a function of the Nadir angle $h$  
and $E/\Delta m^2$, for $\sin^22\theta = 0.01$~(a) and 0.05 (b).

\vspace{0.1cm}
\noindent
{\bf Figure 11.}
The probability $P_{e2}$ as a function of $\rho_r$  
in the case of $\nu_e - \nu_{\mu (\tau)}$ mixing
with $\sin^22\theta = 0.4$. 

\vspace{0.1cm}
\noindent
{\bf Figure 12.}
The probability $P_{e2}$ as a function of $\rho_r$
for $\sin^22\theta = 0.5$  
in the case of $\nu_e - \nu_{s}$ mixing.

\vspace{0.1cm}
\noindent
{\bf Figure 13.} The same as in figure 11 for  
$\sin^22\theta = 0.9$.

\vspace{0.1cm}
\noindent
{\bf Figure 14.} The dependence of the function
$\sin^2(2\theta''_{m} - 4\theta'_{m} + \theta)$ 
on $\rho_r$ for $\sin^22\theta = 0.005;~0.01;~0.05;~0.10~;0.50;~0.90$.

\vspace{0.1cm} 
\noindent
{\bf Figures 15a - 15c.} 
The dependence of the probability $P(\nu_{\mu} \rightarrow \nu_{e}) = 
P(\nu_e \rightarrow \nu_{\mu (\tau)}) = P_{e\mu}$ 
on  $\rho_r$ for 
$h = 0^{\circ}~(a);~ 13^{\circ}~(b);~ 23^{\circ}~(c)$
and $\sin^22\theta = 0.005;~0.01;~0.050;~0.10;~0.50~;0.70$.

\end{document}